\documentclass[sigconf]{acmart}
\makeatletter
\def\@ACM@checkaffil{% Only warnings
    \if@ACM@instpresent\else
    \ClassWarningNoLine{\@classname}{No institution present for an affiliation}%
    \fi
    \if@ACM@citypresent\else
    \ClassWarningNoLine{\@classname}{No city present for an affiliation}%
    \fi
    \if@ACM@countrypresent\else
        \ClassWarningNoLine{\@classname}{No country present for an affiliation}%
    \fi
}
\makeatother

\usepackage{hyperref}
\usepackage{amsmath,amsfonts}
\usepackage{algorithmic}
\usepackage{tikz}
\usepackage{graphicx}
\usepackage{textcomp}
\usepackage{xcolor}
\usepackage{booktabs}
\usepackage{pifont}
\usepackage{siunitx}
\usepackage{amsmath}
\usepackage{algorithm2e}
\usepackage{multirow}
\usepackage{fancyhdr}
\usepackage{float}
\usepackage{setspace}
\usepackage{listings}
\usepackage{balance}
\usepackage{wrapfig}
\usepackage{marginnote}
\usepackage{caption}
\usepackage{boldline} % for bold table borders
\usepackage{colortbl} % for table background color
\usepackage{floatflt}
\usepackage{tabularray}

\usepackage{eso-pic}
\usepackage{titlesec}
\usepackage{noindentafter}
\usepackage{datetime}
\usepackage[htt]{hyphenat}

\AtBeginDocument{%
  \providecommand\BibTeX{{%
    \normalfont B\kern-0.5em{\scshape i\kern-0.25em b}\kern-0.8em\TeX}}}

\usepackage{xspace}
\usepackage{color}
\usepackage{listings}
\usepackage{xcolor}

\definecolor{darkspringgreen}{rgb}{0.09, 0.45, 0.27}
\definecolor{denim}{rgb}{0.08, 0.38, 0.74}
\definecolor{darkolivegreen}{rgb}{0.33, 0.42, 0.18}
\definecolor{tangerine}{rgb}{0.95, 0.52, 0.0}
\definecolor{mahogany}{rgb}{0.75, 0.25, 0.0}

\definecolor{uglyyellow}{rgb}{0.99, 0.93, 0.0}
\definecolor{nbs}{rgb}{0.35, 0.31, 0.81}
\newcommand{\nb}[1]{\textcolor{black}{#1}}

\definecolor{darkred}{rgb}{0.75, 0.1, 0.1}

\newcommand\konkanello[1]{\textcolor{black}{#1}}

\usepackage{tikz}

\newcommand{\squishlist}{
 \begin{list}{$\circ$}
  { \setlength{\itemsep}{0pt}
     \setlength{\parsep}{0pt}
     \setlength{\topsep}{3pt}
     \setlength{\partopsep}{0pt}
     \setlength{\leftmargin}{1em}
     \setlength{\labelwidth}{1em}
     \setlength{\labelsep}{0.5em} } }

\newcommand{\squishend}{
  \end{list}  }

\newcommand*\circled[1]{\tikz[baseline=(char.base)]{\node[shape=circle,fill,inner sep=0.5pt] (char) {\textcolor{white}{#1}};}}

\usepackage[colorinlistoftodos,prependcaption,textsize=scriptsize]{todonotes}

\usepackage[bottom]{footmisc}
\usepackage{dblfloatfix}
\usepackage{array}
 \usepackage{booktabs}

\newcommand\head[1]{{\noindent\textbf{#1}.}}
\definecolor{ufogreen}{rgb}{0.1, 0.6, 0.4}
\definecolor{ufogreen}{rgb}{0.1, 0.6, 0.4}

\newcommand{\papername}{Utopia\xspace}
\newcommand{\speedupsc}{24}
\newcommand{\speedupoverech}{16}
\newcommand{\speedupoverrmm}{11}
\newcommand{\speedupoverpomtlb}{21}
\newcommand{\speedupsecond}{13}

\newcommand{\speedupwithinideal}{95}

\newcommand{\speedupech}{8}
\newcommand{\speeduprmm}{13}

\newcommand{\rowbufferutopia}{20}
\newcommand{\rowbufferech}{50}
\newcommand{\rowbufferptlb}{30}
\newcommand{\ptwdram}{43}

\newcommand{\area}{0.64}
\newcommand{\power}{0.72}

\usepackage{listings}
\usepackage{xcolor}
\usepackage[T1]{fontenc} % Needed for correct font encoding
\usepackage{zi4} % Load the Inconsolata font package
\usepackage{caption} % Load the caption package
\usepackage{float} % Load the float package

\lstdefinestyle{custompseudocode}{
  belowcaptionskip=1\baselineskip,
  breaklines=true,
  xleftmargin=\parindent,
  language=Python,
  showstringspaces=false,
  basicstyle=\small\ttfamily,
  keywordstyle=\bfseries\color{green!40!black},
  commentstyle=\itshape\color{purple},
  stringstyle=\color{orange},
  numbers=left,
  numberstyle=\scriptsize\color{black},
  numbersep=8pt,
  morekeywords={function}, % Added function keyword
  keywordstyle=[2]\bfseries, % Added new keyword class with bold style
  escapeinside={*@}{@*}, % Added escape sequence
  xleftmargin=2em, xrightmargin=0em, 
}
\DeclareCaptionFormat{listing}{% Define the caption format
  \rule{\linewidth}{0.5pt}
  \vspace{-1mm}
  \textbf{#1#2: }#3
  \vspace{-1mm}
  \rule{\linewidth}{0.5pt}
  \vspace{-7mm}
}

\definecolor{cadmiumgreen}{rgb}{0.0, 0.42, 0.24}
\definecolor{chocolate}{rgb}{0.92, 0.41, 0.12}
\definecolor{burgundy}{rgb}{0.5, 0.0, 0.13}
\definecolor{darkmagenta}{rgb}{0.55, 0.0, 0.55}
\definecolor{darkblue}{rgb}{0.0, 0.5, 1.0}
\newcommand\konkanelloreva[1]{\textcolor{black}{#1}}
\newcommand\konkanellorevb[1]{\textcolor{black}{#1}}
\newcommand\konkanellorevc[1]{\textcolor{black}{#1}}

\newcommand\konkanelloreve[1]{\textcolor{black}{#1}}

\newcommand\konreva[1]{\textcolor{black}{#1}}
\newcommand\konrevb[1]{\textcolor{black}{#1}}
\newcommand\konrevc[1]{\textcolor{black}{#1}}
\newcommand\konrevd[1]{\textcolor{black}{#1}}
\newcommand\konrevs[1]{\textcolor{black}{#1}}
\newcommand\konrevf[1]{\textcolor{black}{#1}}
\newcommand\konrevw[1]{\textcolor{black}{#1}}
\newcommand\konrevx[1]{\textcolor{black}{#1}}
\newcommand\konrevl[1]{\textcolor{black}{#1}}

\newcommand\utopiaseg{RestSeg\xspace}
\newcommand\utopiasegs{RestSegs\xspace}
\newcommand\flexseg{FlexSeg\xspace}
\newcommand\flexsegs{FlexSegs\xspace}

\newfloat{listing}{tbhp}{lst}[section] % Create a custom float environment
\floatname{listing}{Listing}
\copyrightyear{2023}
\acmYear{2023}
\setcopyright{acmlicensed}\acmConference[MICRO '23]{56th Annual IEEE/ACM International Symposium on Microarchitecture}{October 28-November 1, 2023}{Toronto, ON, Canada}
\acmBooktitle{56th Annual IEEE/ACM International Symposium on Microarchitecture (MICRO '23), October 28-November 1, 2023, Toronto, ON, Canada}
\acmPrice{15.00}
\acmDOI{10.1145/3613424.3623789}
\acmISBN{979-8-4007-0329-4/23/10}

 \newif\ifcameraready
 \camerareadytrue
\newif\ifarxiv
\arxivtrue

\ifarxiv
\usepackage{calc}
\fancypagestyle{firstpage}
{
    \fancyhead{}
  
  \pagenumbering{arabic}
  \fancyfoot[C]{\large\thepage}
}
\fi

\newcommand\VMcharacterization{~\cite{vm2,karakostas_characterIISWC,
isca2010-barr-trancache,5-levelpaging,contiguitas2023,radiantISMM21,bhattacharjeePACT2009,devirtualizingASPLOS2018,hash_dont_cache,virtualizationimplication,vm29}}

\newcommand\VMlargepages{~\cite{park2020perforated,guvenilir2020tailored,ingensOSDI2016,talluriISCA1992,panwar2018making,panwar2019hawkeye,tridentMICRO2021,pham2015,mosaic2017MICRO,promotionHPCA2001,shadowpageISCA1998,duHPCA2015,vm42,vm43,partialMICRO2020,ganapathy98}}

\newcommand\VMcontiguity{~\cite{translationranger2019,karakostas2015,chloe2020,hybridtlbISCA2017,flexpointerTACO2023,contiguitas2023,vm6,vm2}}

\newcommand\VMtlblthree{~\cite{sharedl3tlbISCA2011,distlltlbMICRO2018,gpustealing}}

\newcommand\VMpwcs{~\cite{isca2010-barr-trancache,vm10,esteve14}}

\newcommand\VMpagetable{~\cite{haria2018, flataAsplos2022,elastic-cuckoo-asplos20,mehtJovanHPCA2023,hash_dont_cache,mosaicpagesASPLOS2023,nearmemoryPact17,impicaICCD2016,mitosis-asplos20,compendiaISMM2021,Alam2017DoItYourselfVM}}

\newcommand\VMvirtualized{~\cite{vm25,vm35,pham2015,pham2015tr,vm11,virtcoherenceISCA2017,babelfish,margaritov2021ptemagnet,virtcoherenceISCA2017,panwar2021fast}}

\newcommand\VMvirtualcaching{~\cite{kaxiras2013,seesawISCA2018,basu2012, cekleov1997a, wood1986,coherencyvirtualASPLOS1987,consistencyvirtualASPLOS1992,virtualcacheISCA1989}}

\newcommand\VMintermediate{~\cite{enigma,midgard,vbi,powerpc2003}}

\newcommand\VMtlbprefetching{~\cite{vavouliotis2021,morriganMICRO2021,margaritov2019prefetched,kandiraju2002going,saulsbury2000recency,Bala1994SoftwarePA}}

\newcommand\VMtlbreplacementpolicy{~\cite{deadTLBHPCA2021,chirpMICRO2020}}

\newcommand\VMtlball{~\cite{chirpMICRO2020,papadopoulou2015,latr,juan97,onlinesuperpagepromotionISCA1995,compilerdtlbISPASS2004,vm38,vm39,wood1986,skewedTLB,vavouliotis2021,morriganMICRO2021,margaritov2019prefetched,kandiraju2002going,saulsbury2000recency,Bala1994SoftwarePA,isca2010-barr-trancache,vm10,sharedl3tlbISCA2011,distlltlbMICRO2018}}

\newcommand\VMsoftwareTLB{~\cite{pomtlbISCA2017,csaltMICRO2017,softwareTLBNAS2013,softwareTLBISCA2013,uhlig94,bruceMMU1998,softcontrolcachesISCA1986,Nagle1993DesignTF,Bala1994SoftwarePA}}

\newcommand\VMold{~\cite{ieemicro2018-Bhattacharjee-tempo,hand1999,old_vm1,old_vm2,old_vm3,old_vm4,old_vm5,old_vm6,old_vm7,denning1970,ahearn1973,goldberg1974survey,bruceMMU1998,smith,wood1986,chen1992simulation,koldinger1992,lindstrom1995,jacob1998,avm,translationmanagementISCA1993,interactionASPLOS1991,multics}}

\begin{document}

 \title{ Utopia: Fast and Efficient Address Translation via Hybrid\\   Restrictive \& Flexible Virtual-to-Physical Address Mappings}
\author{
  Konstantinos Kanellopoulos\textsuperscript{1}\quad
  Rahul Bera\textsuperscript{1}\quad
  Kosta Stojiljkovic\textsuperscript{1}\quad 
  Nisa Bostanci\textsuperscript{1}\quad 
  Can Firtina\textsuperscript{1}\quad 
  Rachata Ausavarungnirun\textsuperscript{2}\quad
  Rakesh Kumar\textsuperscript{3}\quad
  Nastaran Hajinazar\textsuperscript{4}\quad
  Mohammad Sadrosadati\textsuperscript{1}\quad
  Nandita Vijaykumar\textsuperscript{5}\quad
  Onur Mutlu\textsuperscript{1}\quad
}
\affiliation{%
  \vspace{0.7em}
  \institution{\textsuperscript{1}ETH Zürich\quad\textsuperscript{2}King Mongkut's University of Technology North Bangkok\quad\\\textsuperscript{3}Norwegian University of Science and Technology\quad\textsuperscript{4}Intel Labs\quad\textsuperscript{5}University of Toronto}
   % \country{Switzerland}
}
\email{}

 \renewcommand{\shortauthors}{Kanellopoulos et al.}
 \renewcommand{\shorttitle}{Utopia: Fast and Efficient Address Translation via Hybrid\\ Restrictive \& Flexible Virtual-to-Physical Address Mappings}

\renewcommand{\authors}{Konstantinos Kanellopoulos,
Rahul Bera,
Kosta Stojiljkovic, 
Nisa Bostanci,
Can Firtina, 
Rachata Ausavarungnirun,
Rakesh Kumar,
Nastaran Hajinazar,
Mohammad Sadrosadati,
Nandita Vijaykumar,
Onur Mutlu}

\begin{abstract}
Conventional virtual memory (VM) frameworks enable a virtual address to flexibly map to \emph{any} physical address. 
This flexibility necessitates large data structures to store virtual-to-physical mappings, which leads to high address translation latency 
and \konreva{large} translation-induced interference in the memory hierarchy, especially in data-intensive workloads.
On the other hand, restricting the address mapping so that a virtual address \konreva{can} only map \konreva{to} a specific set of physical addresses can significantly 
reduce address translation overheads by making use of compact and efficient translation structures.
However, restricting the address mapping flexibility across the entire main memory
severely limits data sharing across different processes and increases \konrevb{data accesses} to the swap space of the storage device even in the presence of free memory. 
%pages might not be able to map to that free memory due to the restrictions imposed by the mapping scheme.

We propose \emph{\papername}, a new hybrid virtual-to-physical address mapping scheme that allows \emph{both} flexible and restrictive hash-based address mapping schemes 
to harmoniously \emph{co-exist} in the system. The key idea of Utopia is to manage physical memory using two types of physical memory segments: restrictive segments and flexible segments.
A restrictive segment uses a restrictive, hash-based address mapping scheme
that maps virtual addresses to \konreva{only} a specific set of physical addresses 
and enables faster address translation using compact translation structures.
A flexible segment employs the conventional fully-flexible address mapping scheme.
By mapping data to a restrictive segment, \papername enables faster address translation with lower translation-induced interference.
At the same time, Utopia retains the ability to use the flexible address mapping to (i) support conventional VM features such as data sharing and 
(ii) \konreva{avoid storing data in the swap space of the storage device} when \konreva{program} data does not fit inside a restrictive segment.

Our evaluation using 11 diverse data-intensive workloads shows that \papername improves performance by \speedupsc\% in \konreva{a single-core system} over the 
baseline \konreva{conventional} four-level radix-tree page table design, whereas the best prior state-of-the-art contiguity-aware translation scheme improves performance by \speedupsecond\%.
\papername provides \speedupwithinideal\% of the performance benefits of an ideal address translation scheme where every translation request hits in the first-level TLB.
All of Utopia's benefits come at a modest cost of \area\% area overhead and \power\% power overhead compared to a modern high-end CPU. The source code of Utopia is freely available
at \textcolor{blue}{\url{https://github.com/CMU-SAFARI/Utopia}}.
\end{abstract}

\begin{CCSXML}
<ccs2012>
<concept>
<concept_id>10010583.10010786.10010809</concept_id>
<concept_desc>Hardware~Memory and dense storage</concept_desc>
<concept_significance>300</concept_significance>
</concept>
<concept>
<concept_id>10011007.10010940.10010941.10010949.10010950.10010951</concept_id>
<concept_desc>Software and its engineering~Virtual memory</concept_desc>
<concept_significance>500</concept_significance>
</concept>
</ccs2012>
\end{CCSXML}

\ccsdesc[300]{Hardware~Memory and dense storage}
\ccsdesc[500]{Software and its engineering~Virtual memory}
%
% Keywords. The author(s) should pick words that accurately describe
% the work being presented. Separate the keywords with commas.
\keywords{Virtual Memory, Cache, TLB, Virtualization, Microarchitecture, \\ Address Translation, Memory Systems}

\maketitle
\thispagestyle{firstpage}

\section{Introduction}

Virtual memory (VM) serves as a foundational element in \konreva{most} computing systems, 
simplifying the programming model by offering an abstraction layer over physical memory\VMold. In the presence of VM, the operating system (OS) maps each virtual address to its corresponding physical memory address to facilitate application-transparent memory management, process isolation, and memory protection.
The virtual-to-physical mapping scheme in conventional VM frameworks allows a virtual address to flexibly map to \emph{any} physical address. 
This flexibility enables key VM functionalities, such as (i) data sharing between processes while maintaining process isolation and (ii) avoiding frequent \konreva{swapping
 (i.e., avoiding storing data in the swap space of the storage device in the presence of free main memory space).}
However, a flexible mapping scheme \konreva{requires} mapping \konreva{metadata for} every virtual address and its corresponding physical address, which is stored in the page table (PT).
As shown in multiple prior works\VMcharacterization, data-intensive workloads do not efficiently use translation-dedicated 
hardware structures and the processor performs frequent PT accesses, i.e., a process called PT walk (PTW), to resolve address translation requests. 
Frequent accesses to the PT heavily impact system performance in two ways: \konreva{they lead to} 
(i) high address translation latency and (ii) interference between program data and the PT data across the memory hierarchy, i.e., CPU caches, \konreva{interconnect} and main memory.

\textbf{High address translation latency.} 
As data-intensive applications use increasingly larger data sets, the size of the PT grows,
which increases the latency of PTWs. For example, modern x86-64 systems use a four-level radix-tree PT that requires up to four \emph{serialized} memory accesses, 
to translate a virtual address to its corresponding physical address~\cite{intelx86manual}. \konreva{For} workloads \konreva{that} make scarce use of the main memory \konreva{capacity},
walking the four-level radix table is fast since the PTs are small enough to fit in on-chip caches.
However, the large data footprints of emerging data-intensive workloads (e.g., graph analytics~\cite{graphAnalyticsSurevey,graph500}, recommendation systems~\cite{dlmr,recommndationSurvey}, generative models~\cite{openai2023gpt4,vaswani2023attention}) 
lead to large PTs that do not fit in on-chip caches. For example, given an application with a 2TB dataset,
the x86-64 PT's size can reach up to 4GB, which is much larger than the total caching capacity of a modern high-end CPU~\cite{raptor_lake}.
As we demonstrate in \S\ref{sec:motivation}, even using the state-of-the-art hash-based PT design~\cite{elastic-cuckoo-asplos20} in a system that supports both 4KB and 2MB pages~\cite{arcangeli2010,corbet2011,reserve,hugepage}, 
\konrevb{a} PTW takes \konreva{an average of 86 cycles (up to 123) to complete}, across 11 data-intensive workloads. \konreva{High frequency} and high latency PTWs lead to high address translation latency and degrade system performance.

\textbf{Translation-induced interference in the memory hierarchy.} During a PTW, the processor issues memory requests to the memory hierarchy in order to fetch the PT. 
Upon retrieval from main memory, the PT data is stored within the cache hierarchy.
As a result, PTWs interfere with the memory hierarchy of a processor in two major ways. First, PTWs consume the scarce on-chip cache hierarchy space \konreva{(to store translation metadata)}, which otherwise could \konreva{be used} to cache program data.
Second, PTWs increase DRAM row buffer misses due to frequent DRAM accesses for retrieving translation metadata. 
In \S\ref{sec:motivation}, we show that (i)  data-intensive applications consume up to $38\%$ of the L2 cache capacity only to store PT data, and 
(ii) memory requests for PT data increase DRAM row buffer misses by 30\% compared to an ideal system that uses a perfect translation-lookaside buffer (TLB).\footnote{i.e., \konreva{a system where} every translation request hits in the L1 TLB.}

Prior works~\cite{nearmemoryPact17,smith,mosaicpagesASPLOS2023} explore the possibility of restricting the virtual-to-physical mapping (\konreva{e.g.}, the physical address can be computed based on a hash function applied to the virtual address) flexibility to reduce the size of 
the data structures that store translation metadata and reduce the address translation overhead. Restricting the virtual-to-physical mapping drastically reduces the size 
of the translation data structures, and accordingly lowers the latency of retrieving the virtual-to-physical mapping. For example, as shown in~\cite{nearmemoryPact17},
determining the physical location of a virtual page based on a specific set of bits of the virtual address is considerably faster than accessing the x86-64 multi-level PT. 
However, \konreva{restricting} the address mapping \emph{across the entire physical address space} in general-purpose systems {handicaps} core
VM functionalities and \konreva{can cause} severe performance overheads. 
First, \konreva{two} virtual pages from different processes might not be able to map to the same physical page which limits data sharing. Second, 
the sole use of a restrictive mapping leads to memory underutilization as the system might not be able to freely map virtual pages to the available free physical space. 
\konreva{This can cause more pages to be stored inside the swap space of the storage device even in the presence of free physical memory space.}
Our analysis in \S\ref{sec:motivation} shows that restricting the flexibility of the address mapping, as proposed in~\cite{nearmemoryPact17}, for the whole main memory of a general-purpose 
system \konrevb{increases data accesses} to the swap space of the storage device by 122\%.
We conclude that using only flexible or only restrictive hash-based address mapping does not satisfy 
the requirements of both highly-flexible memory management and high-performance, low-interference address translation.  

\textbf{Our goal} is to design a virtual-to-physical address mapping scheme that provides fast and efficient translation \konreva{via the use of a} restrictive hash-based address mapping while still enjoying the benefits of the conventional fully-flexible address mapping. 
To this end, we propose \emph{Utopia}, a new hybrid virtual-to-physical address mapping scheme that enables \emph{both} flexible and  restrictive hash-based virtual-to-physical address mapping schemes to harmoniously \emph{co-exist} in the system.
%in order to mitigate the overheads associated with address translation, while maintaining the benefits of the flexible address mapping.
\textbf{The key idea} of Utopia is to manage physical memory using two types of physical memory segments: \emph{restrictive} segments and the \emph{flexible} segment.
%in a hybrid manner
A restrictive segment (called \emph{\utopiaseg}) enforces a restrictive, hash-based address mapping 
scheme, \konreva{thereby} enabling fast and efficient address translation through compact and efficient address translation structures.
A flexible segment (called \emph{\flexseg}), employs the conventional address mapping scheme and provides full virtual-to-physical address mapping flexibility. 
By mapping data to a \utopiaseg, Utopia enables fast address translation with low translation-induced interference in the memory hierarchy 
%{memory interference} 
whenever flexible address mapping is not necessary (e.g., \konreva{when optimizing for fast address translation}). At the same time, Utopia retains the ability to use flexible address mapping to (i) support conventional virtual memory features such as data sharing and 
(ii) avoid accesses \konreva{to the swap space} when data \konreva{does} not fit inside a restrictive segment.

%\todo{@Kanell do we agree with the use of "hash-based" mapping?}
\textbf{Key Mechanism.}
\konreva{We} study an example implementation of Utopia that uses a set-associative address mapping (similar to how hardware caches work) as the restrictive hash-based address mapping scheme to map data to \utopiasegs. 
In contrast to the conventional flexible address mapping, which requires expensive PTWs to resolve a translation request, the set-associative address mapping scheme  requires \konreva{only}
(i) calculating the set index by applying a hash function over the virtual address, and (ii) performing tag matching using a highly-compact and scalable data structure compared to the conventional PT.

\textbf{Key Challenges.}
Integrating Utopia into a conventional system requires addressing three key challenges. 
First, identifying the pages that are good candidates for storing inside a \utopiaseg. To address this challenge, we propose a mechanism to predict whether \konreva{or not}
a page is \konreva{costly to translate}. Based on this prediction, Utopia determines the potential benefits of allocating \konreva{the page} within a \utopiaseg (\S\ref{sec:heuristics}).
Second, efficiently managing the co-existence of \utopiasegs and \flexsegs in a single physical address space. 
We address this challenge by extending the \konreva{operating system} to support the creation of \utopiasegs, \konreva{allocation} of pages in \utopiasegs and \flexsegs, and 
\konreva{migration of} pages between the two \konreva{segments} (\S\ref{sec:os-support}). Third, accelerating address translation for pages that reside in a \utopiaseg. We achieve this by extending the 
processor with architectural support to efficiently access the data structures of the \utopiaseg with minimal 
overhead on top of the existing PTW path (\S\ref{sec:arch-support}).

\textbf{Key Results.}
We evaluate \papername with an extended version of the Sniper simulator~\cite{sniper} (which \konreva{we} open-source ~\cite{utopia_github}) using 11 data-intensive applications from five diverse benchmark suites 
(GraphBIG~\cite{Lifeng2015}, GUPS~\cite{Plimpton2006}, XSBench~\cite{Tramm2014}, DLRM~\cite{dlmr} and GenomicsBench~\cite{genomicsbench}). Our evaluation yields five key results that \konreva{demonstrate} Utopia's effectiveness.
First, in single-core (four-core) workloads, Utopia improves performance by \speedupsc\% (28\%) on average over the \konreva{conventional} four-level radix-tree \konreva{baseline} PT design, whereas two prior state-of-the-art translation mechanisms, elastic cuckoo hashing (ECH~\cite{elastic-cuckoo-asplos20}) and redundant memory mappings (RMM~\cite{karakostas2015}) improve performance by \speedupech\% (14\%) and \speeduprmm\% (12\%), respectively.
Second, in single-core workloads, Utopia provides 95\% of the performance of an ideal \konreva{address} translation scheme where every translation request hits in the L1 TLB.
Third, Utopia reduces \konreva{address translation} latency by 69\% over the baseline radix-based page table in single-core workloads, whereas ECH and RMM reduce translation latency by 39\% and 15\%, respectively.
Fourth, Utopia reduces DRAM row buffer misses by \rowbufferutopia\% compared to the baseline system.
Fifth, all of Utopia's benefits come at a modest cost of $0.74\%$ area overhead and $0.62\%$ power overhead compared to a modern high-end CPU~\cite{raptor_lake}.

\noindent We make the following major contributions in this paper:

\begin{itemize}
    \item We \konrevb{demonstrate} that \konreva{although} using a restrictive virtual-to-physical address mapping reduces 
    the address translation overhead and memory hierarchy interference, 
    employing a restrictive address mapping across the \emph{entire} physical address space limits key benefits of VM design 
    such as data sharing between processes and high memory utilization.

    \item We propose \emph{Utopia}, a new hybrid virtual-to-physical address mapping scheme that allows both  \emph{flexible}  
    and \emph{restrictive} hash-based virtual-to-physical address mapping schemes to harmoniously co-exist in the system. 
    This way, Utopia enables efficient address translation and reduces translation-induced interference in the memory hierarchy 
    while maintaining the benefits of a fully-flexible address mapping scheme.
    
    \item We devise three key components to integrate Utopia into a conventional system: 
    (i) a set of techniques to identify costly-to-translate pages \konreva{that} benefit from a restrictive mapping, 
    (ii) OS support to manage the co-existence of flexible and restrictive address mappings in the same physical address space, and 
    (iii) lightweight architectural support to integrate Utopia in the processor's translation pipeline.
    
    \item We quantitatively evaluate Utopia in single-core and multi-core environments and compare it against three state-of-the-art address translation mechanisms. 
    Our experimental results show that Utopia significantly reduces the overheads associated with address translation at a modest low area and power cost. 
    We open-source Utopia at \textcolor{blue}{\url{https://github.com/CMU-SAFARI/Utopia}}.

\end{itemize}

\section{Background}
\label{sec:background}
%\outline{
%\begin{itemize}
%    \item What is virtual memory? 
%    \item We have fully associative main memory: how do we store the mapping info? Page table. 
%    \item How is page table organized? 4-level page radix tree. We need 4 memory accesses to resolve the translation request
%    \item How does the system take care of address translation? Memory Management Unit: what does it handle? (i) caching %mappings in the TLB (ii) handling permissions and (iii) specialized hardware to perform the page table walk.
%    \item TLBs are not able to handle memory hungry workloads: Figure with results
%    \item TLB misses result in expensive page table walks that dominate execution time: Figure with results. 
%\end{itemize}
%}
\subsection{The Virtual Memory Abstraction}
Virtual memory is a cornerstone of most modern computing systems
\konrevb{that} eases the programming model by providing a \konreva{convenient} abstraction to the physical memory\VMold. The operating system (OS), transparently \konreva{to application software}, maps each virtual memory address to its corresponding physical memory address. \konreva{Doing so} 
{provides a number of benefits, including: (i) application-transparent memory management, (ii) sharing data between applications, (iii) process isolation and (iv) \konreva{page-level} memory protection.}
Conventional virtual memory designs allow any virtual page to map to any free physical page. Such a flexible address mapping 
enables two important key features of virtual memory: (i) efficient memory utilization to avoid frequent swapping,\footnote{Swap space is a reserved space in the storage device that is used to store pages that are not currently mapped in the main memory~\cite{swapspace,swapspace2,OSF}. 
When a page is evicted from the main memory, the OS stores it in the swap space. When the page is accessed again, the OS loads it back into the main memory.} and (ii) sharing pages between applications.
However, \konreva{such a} flexible address mapping \konreva{mechanism} has a critical downside: it creates the need to store a large number of virtual-to-physical mappings, \konrevb{since the OS needs to keep track of the physical
location of \konrevd{every} virtual page \konrevd{that is} used by each process}.

\subsection{Page Table (PT)}

{The \konreva{PT} is a per-process data structure that stores the mappings between virtual and physical pages. 
In modern x86-64 processors, the \konreva{PT} is organized as a four-level radix-tree~\cite{intelx86manual}. Even though the radix-tree-based PT 
optimizes for storage efficiency, it requires \konreva{multiple} pointer-chasing operations to discover the virtual-to-physical mapping.  
To search for a virtual-to-physical address mapping, the system needs to \emph{sequentially} access each of the four-levels of the page table. 
This process is called \konrevb{a} \emph{page table walk (PTW)}.}

\begin{figure}[h]
    \vspace{-2mm}
    \centering
    \includegraphics[width=\columnwidth]{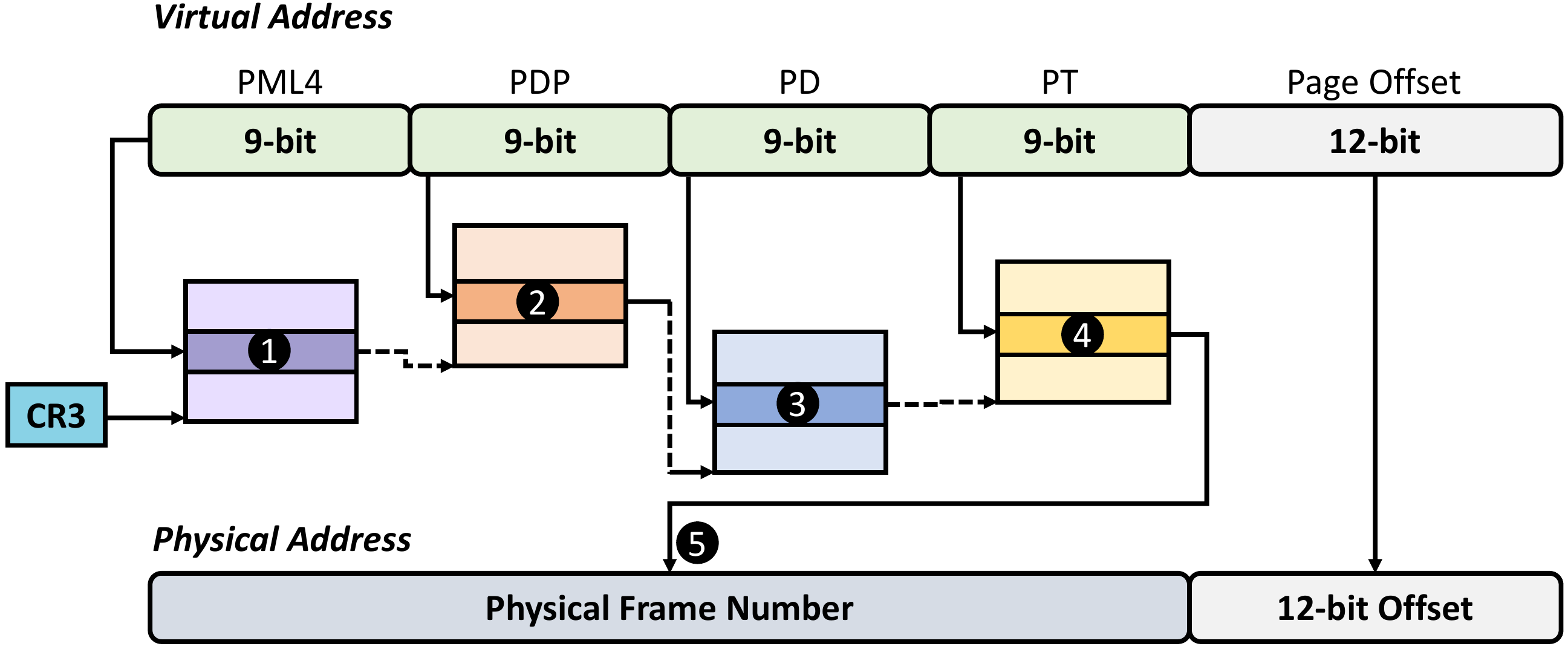}
    \vspace{-7mm}
    \caption{Four-level radix-tree page table walk in x86-64 ISA.}
    \label{fig:page-table}
    \vspace{-2mm}
\end{figure}

{Figure~\ref{fig:page-table} shows the page table walk assuming (i) an x86-64 four-level radix-tree page table whose base address is stored in the CR3 register,\footnote{\konrevb{CR3 register stores the page table base address in the x86-64 ISA.}} and (ii) 4KB pages.
 As shown in Fig.~\ref{fig:page-table}, a single PTW requires four sequential memory accesses \konreva{\circled{1}-\circled{4}} to discover the physical page number. 
 The processor uses the first 9-bits of the \konrevb{48-bit} virtual address as offset \konreva{(Page Map Level4; PML4)} to index the appropriate entry of the page table within the first level of the page table~\circled{1}. 
 The processor then reads the pointer stored in the first level of the page table to access the second-level of the page table~\circled{2}. 
 It uses the next 9-bit set \konreva{(Page Directory Page table; PDP)} from the virtual address to locate the appropriate entry within the second level. 
 This process continues \konreva{iteratively} for each subsequent level of the \konrevb{multi-level (hierarchical)} page table \konreva{(Page Directory; PD~\circled{3} and Page Table; PT~\circled{4})}.
Eventually, the processor reaches the leaf level of the page table, where it finds the final entry containing the physical page number corresponding to the given virtual address~\circled{5}. 
As shown in multiple previous works\VMcharacterization, PTWs incur high translation latency, which leads to high performance overheads. ARM processors use a similar approach, with the number of levels varying across different versions of the ISA~\cite{arm-manual-tlbmaintenance}.}

\subsection{Memory \konreva{Management} Unit (MMU)}
\begin{figure}[h]
    \vspace{-2mm}
    \centering
    \includegraphics[width=3.3in]{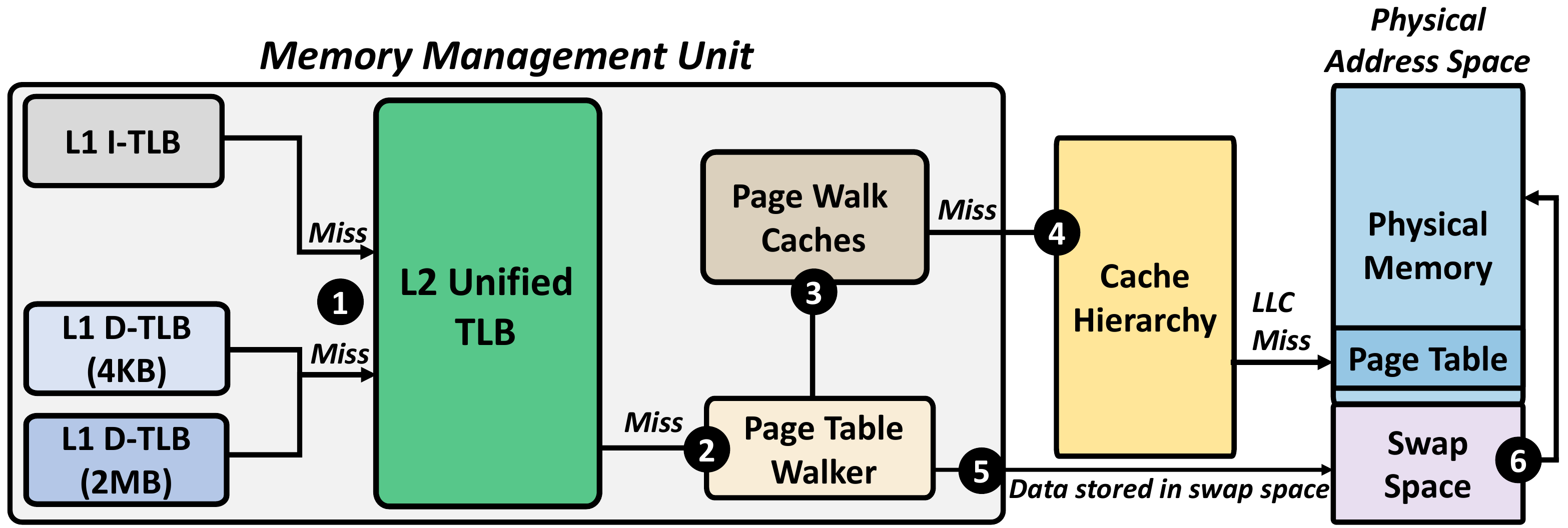}
    \vspace{-2mm}
    \caption{Structure of the Memory Management Unit (MMU) of a modern processor.}
    \label{fig:mmu}
    \vspace{-2mm}

\end{figure}

\konreva{When} a user process {generates} a memory (i.e., instruction or data) request, the processor needs to translate the virtual address to its corresponding physical address. Address translation is a critical operation because \konrevb{it is} on the critical path of the memory access flow: no memory access is possible unless the requested virtual address is first translated into its corresponding physical address.
Given that frequent page \nb{table} walks lead to high address translation overheads, modern cores employ a specialized memory management unit (MMU) responsible for accelerating address translation.
\autoref{fig:mmu} {shows an example structure of the MMU of a modern processor~\cite{cascadelake}, consisting of three key components: (i) a two-level hierarchy of translation lookaside buffers (TLBs), (ii) a hardware page table walker, (iii) page walk caches (PWCs).}

L1 TLBs are {highly- or fully-}associative \konrevb{translation} caches that directly provide the physical address for recently-accessed virtual pages at very low latency (i.e., typically within 1 cycle).
There are two separate L1 TLBs, one for instructions (L1 I-TLB) and one for data (L1 D-TLB). 
{Modern TLBs make use of multiple page sizes beyond 4KB in order to (i) cover large amounts memory with a single \konrevb{PT} entry and (ii)  maintain compatibility with modern OSes that transparently allocate large pages~\cite{tridentMICRO2021,corbet2011,reserve,panwar2019hawkeye}. 
For example, \konreva{an Intel} Cascade Lake core~\cite{cascadelake} employs 2 L1 D-TLBs, \konreva{one for 2MB and one for 4KB pages.}}
Translation requests that miss in the L1 TLBs~\circled{1} are forwarded to a unified L2 TLB that stores translations for both instructions and data. 
In case of an L2 TLB miss, the MMU triggers a PTW~\circled{2}. 
PTW is performed by a dedicated hardware page table walker capable of performing multiple concurrent PTWs.
In order to reduce PTW latency, page table walkers are equipped with page walk caches (PWC)~\circled{3}, which are small dedicated caches for each level of the PT (\konrevb{e.g.,} for the first three levels in x86-64). 
In case of a PWC miss, the MMU issues the request(s) for the corresponding level of the PT to the conventional memory hierarchy~\circled{4}.
\konrevd{If the physical address points to a page inside the swap space~\cite{linux-swap} of the storage device~\circled{5}, the MMU issues a request to the storage device to move the page from the swap space into the main memory~\circled{6}.
If the physical address is not found in the PT, the MMU raises a page fault exception to pass control to the OS.}

%This results in the system having to perform up to four memory accesses for every address translation request, as demonstrated in the Figure \ref{fig:page-table}.

% \begin{figure}[b]
%     %\vspace{-3mm}
%     \centering
%     \includegraphics[width=3.3in]{figures/motivation/memory_management_unit.pdf}
%     %\vspace{-1mm}
%     \caption{\js{Overview of} Memory Management Unit\js{.}}
%     \label{fig:mmu}
%     %\vspace{-3mm}
% \end{figure}

\section{Motivation}

\label{sec:motivation}

\begin{figure}[h!]
    \vspace{-2mm}
    \centering
    \includegraphics[width=\columnwidth]{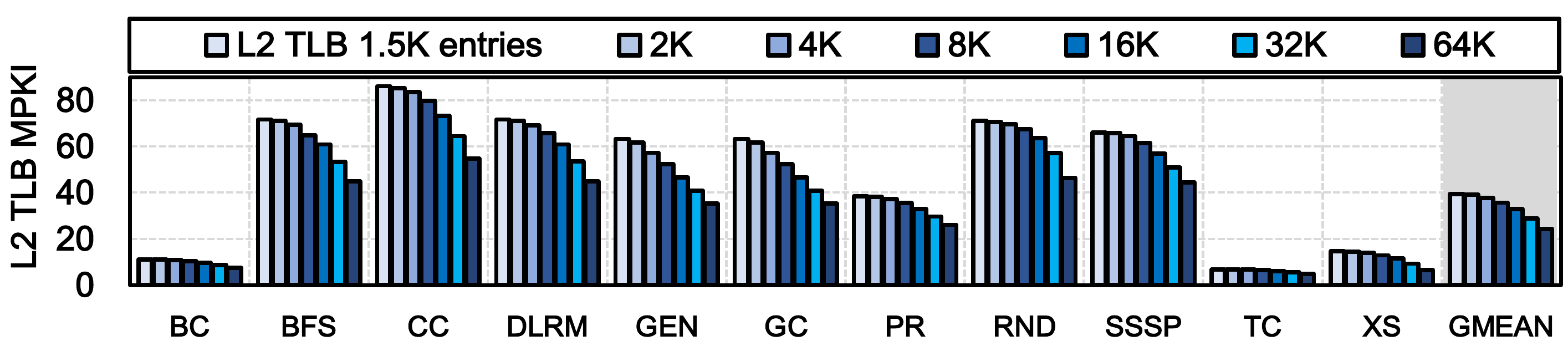}
    \vspace{-7mm}
    \caption{{L2 TLB MPKI across L2 TLBs with different sizes.}}
    \label{fig:l2tlb_mpki}
    \vspace{-2mm}
\end{figure}

{
Data-intensive workloads (e.g., graph analytics~\cite{graphAnalyticsSurevey,graph500}, recommendation systems~\cite{dlmr,recommndationSurvey}, generative models~\cite{openai2023gpt4,vaswani2023attention})
use large datasets and exhibit irregular memory access patterns that lead to
large and costly-to-access page tables. Multiple prior works\VMcharacterization~and large-scale industrial studies~\cite{radiantISMM21,contiguitas2023} demonstrate that
a wide range of data-intensive workloads experience high TLB miss rates and high PTW latencies. 
Figure \ref{fig:l2tlb_mpki} shows the L2 TLB MPKI of the baseline system, as we increase the L2 TLB size from 1.5K entries up to 64K entries, for 11 memory-intensive workloads.\footnote{Section~\ref{sec:methodology} describes in detail our evaluation methodology.}
We observe that the baseline 1.5K-entry L2 TLB suffers from high \konrevb{average} MPKI, 39 on average and up to 77. 
Even using a drastically larger 64K-entry L2 TLB, the \konrevb{average} MPKI remains high \konrevb{at 24} \konrevb{(and up to 54)}, resulting in frequent PTWs.
Frequent and high-latency PTWs pose two key challenges that significantly impact system performance: 
(i) high address translation latency and (ii) \konrevb{high} address translation-induced interference in the memory hierarchy. 
}

\begin{figure}[h!]
    \vspace{-2mm}
    \centering
    \includegraphics[width=\columnwidth]{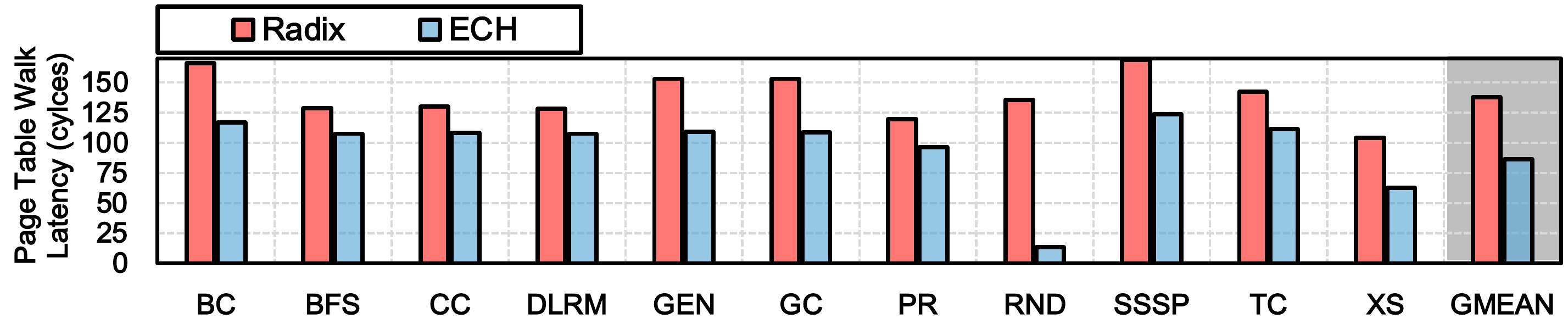}
    \vspace{-7mm}
    \caption{Average PTW latency in Radix and ECH.}
    \label{fig:avgptwlat}
    \vspace{-3mm}
\end{figure}

\head{High Address Translation Latency}  To better understand the performance of address translation in data-intensive workloads, 
we study the effectiveness of two different systems: (i) a baseline system that uses the conventional four-level radix page table (Radix) and  (ii) a system that uses the state-of-the-art 
\konrevb{elastic cuckoo hash-based page table (ECH)} proposed in~\cite{elastic-cuckoo-asplos20}. Both systems use 4KB and 2MB pages allocated by \konrevb{the Linux Transparent Huge Page (THP) mechanism}~\cite{arcangeli2010,corbet2011}.  
Figure~\ref{fig:avgptwlat} shows the average PTW latency (in processor cycles) for Radix and ECH. 
We observe that Radix spends 137 cycles and ECH 86 cycles, \konrevb{on average,} to complete the PTW. 

\begin{figure}[h!]
    \vspace{-2mm}
    \centering
    \includegraphics[width=\columnwidth]{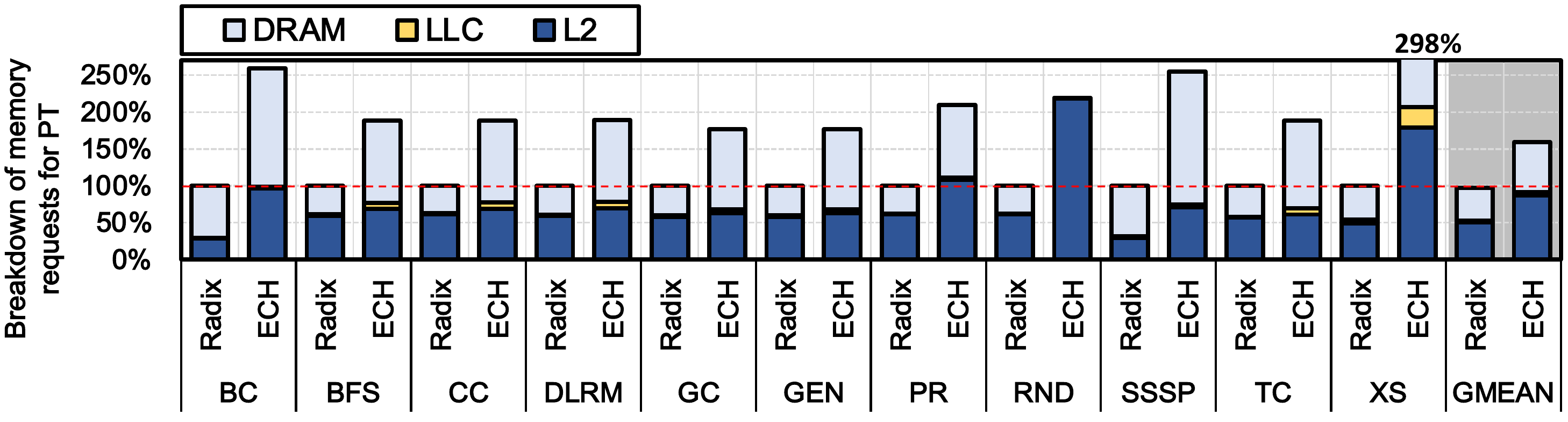}
    \vspace{-7mm}
    \caption{\konrevf{Breakdown of the \konrevb{servicing} location of memory requests issued during PTWs for ECH normalized to Radix.}}
    \label{fig:pt_cache_hit_rate}
    \vspace{-2mm}
\end{figure}

Figure~\ref{fig:pt_cache_hit_rate} demonstrates the breakdown of the \konrevb{servicing} location (DRAM, LLC, L2) of memory requests to access the PT in \konrevb{both Radix and ECH, normalized to Radix}. We make two key observations. 
First, \konrevb{an average of} \ptwdram\% of the PT requests are \konrevb{serviced} from DRAM, \konrevb{in Radix}. This is the key reason behind the long average PTW latency \nb{of Radix} (137 cycles). 
Second, \konrevb{although} ECH reduces the fraction of PT requests that hit in the DRAM, it increases the \emph{total} number of memory requests (to access the PT) by $62\%$ on average compared \nb{Radix}. 
This is because ECH looks up \konrevb{4} hash tables in parallel and issues multiple memory requests to the memory hierarchy. 
\konrevb{Yet}, only one of the issued requests \konrevb{is actually necessary} \konrevb{(i.e. one request will fetch the correct virtual-to-physical address translation)}.\footnote{\konrevb{ECH issues more memory requests than Radix as (i) it employs $n=4$ hash tables and accesses them in parallel and (ii) the 
entries of the hash tables are not cached in a specialized component similar to the PWC in Radix. Figure~\ref{fig:avgptwlat} reports the PTW latency \emph{only} for the requests that deliver the translation.
Figure~\ref{fig:pt_cache_hit_rate} reports the servicing location of all the memory requests issued by the page table walker.}}
We conclude that \konrevb{although} (i) ECH reduces the average PTW latency compared to Radix and (ii) \konrevb{our evaluated} system employs both 4KB and 2MB pages, the 
average PTW latency still remains high.

\begin{figure}[h!]
    % \vspace{-2mm}
    \centering
    \includegraphics[width=\columnwidth]{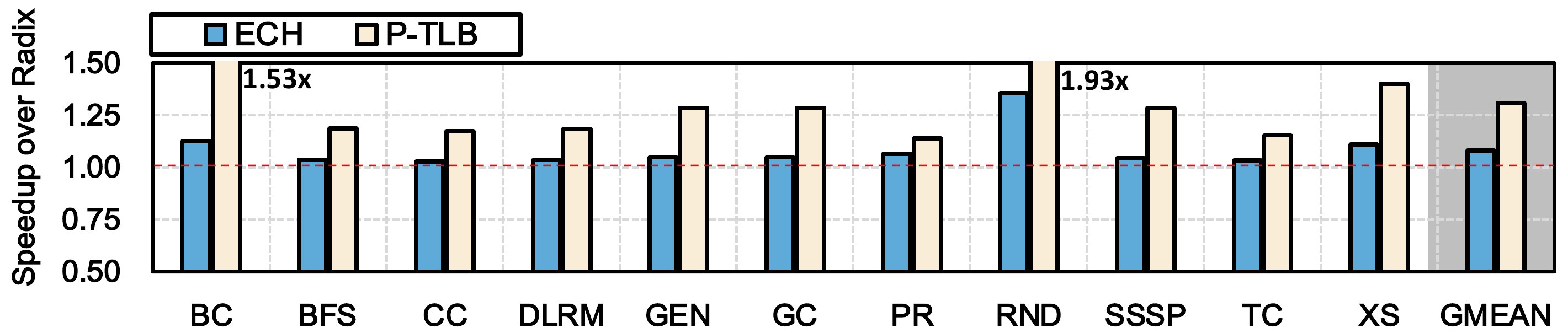}
    \vspace{-7mm}
    \caption{Speedup of ECH and \konrevb{perfect TLB} over Radix.}
    \label{fig:llc_mpki}
    \vspace{-2mm}
\end{figure}

To better understand the headroom for improving the performance of address translation, we evaluate the performance of an ideal system that employs a perfect L1 TLB (P-TLB).\footnote{\konrevb{Every translation request hits in the L1 TLB.}}
Figure~\ref{fig:llc_mpki} shows the \konrevb{execution time} speedup of ECH and P-TLB compared to Radix. 
\konrevb{We observe that} P-TLB outperforms Radix by $30\%$ and ECH by $22\%$. 
We conclude that there is room for \konrevb{further} improving the performance of address translation.

\begin{figure}[h!]
    \vspace{-2mm}
    \centering
    \includegraphics[width=\columnwidth]{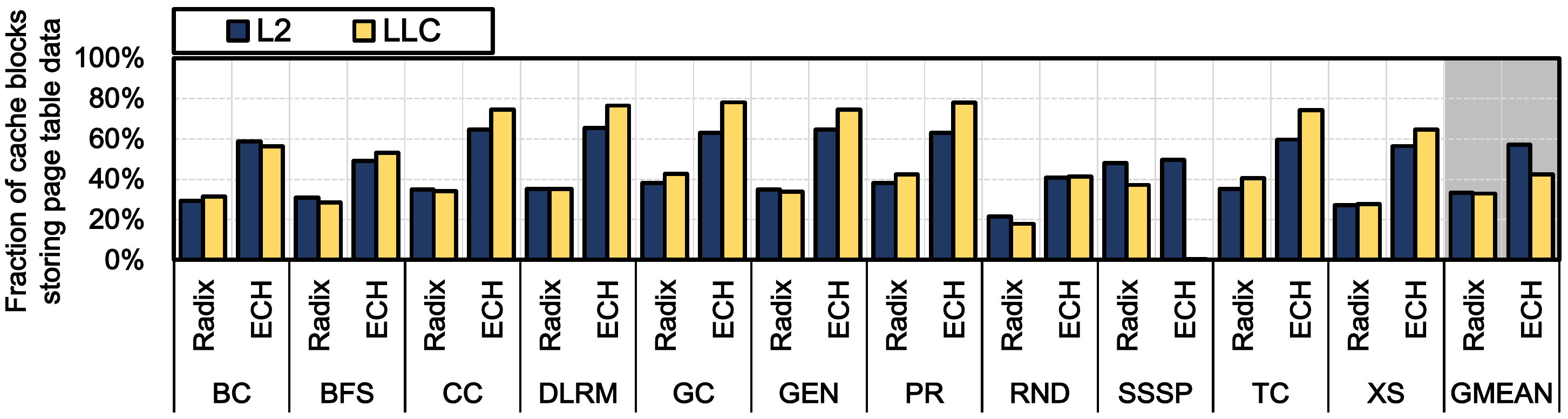}
    \vspace{-7mm}
    \caption{Fraction of \konrevb{cache} blocks that contain PT entries.}
    \label{fig:capacity}
    \vspace{-2mm}

\end{figure}

\head{Translation-induced Interference in Memory Hierarchy}
To better understand the impact of PTWs on the system, we evaluate \konrevb{two example measures of memory interference:} 
(i) \konrevb{the \konrevd{fraction} of cache blocks that store PT entries across the cache hierarchy}, and 
(ii) how address translation affects DRAM row buffer misses.
Figure~\ref{fig:capacity} shows the fraction of cache blocks of two caches (L2, LLC) that store PT data (L1 typically does not store PT entries~\cite{flataAsplos2022,consciousISPASS2022,pinningAccess2022}), averaged across 500 \nb{epochs of} 1M instructions, for Radix and ECH. 
\konrevb{We observe that} both Radix and ECH \konrevb{use} significant \konrevb{fraction} of cache capacity in the cache hierarchy. For example, Radix and ECH respectively \konrevb{use} 33\% and 57\% of L2's total capacity for PT entries.
The high \konrevb{usage} of cache blocks for PT entries reduces the effective capacity of the cache hierarchy, which otherwise could have been used to store \konrevb{the data of} (i) the running application and (ii) other applications running on the system \konrevb{if the LLC is shared.}

\begin{figure}[h!]
    \vspace{-2mm}
    \centering
    \includegraphics[width=\columnwidth]{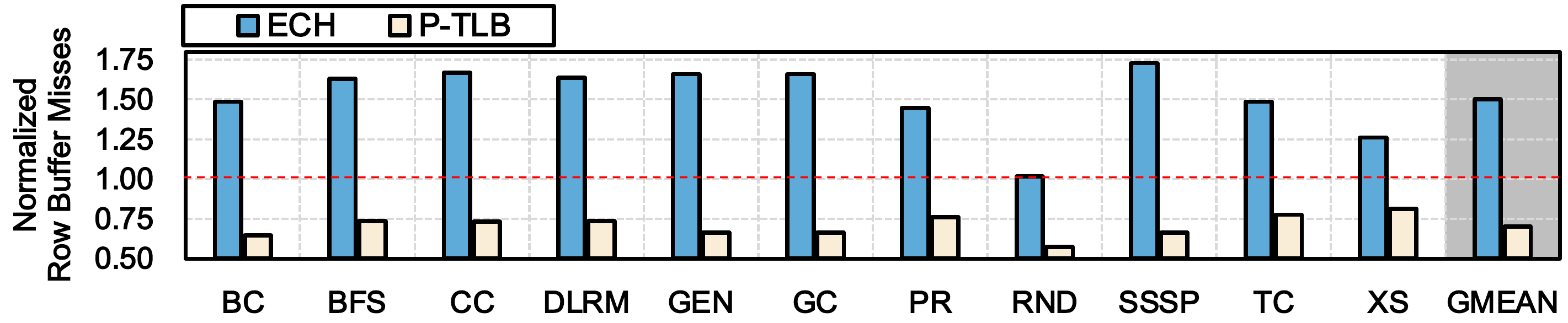}
    \vspace{-7mm}
    \caption{Normalized DRAM row buffer conflicts for ECH and \konrevb{perfect TLB} over Radix.}
    \label{fig:row-buffer}
    \vspace{-2mm}
\end{figure}

Figure~\ref{fig:row-buffer} shows the reduction \konrevb{in} DRAM row buffer conflicts \konrevb{provided by} ECH and a perfect L1 TLB (P-TLB) compared to Radix. 
\konrevb{We observe that} \konrevb{ (i) ECH increases DRAM row buffer conflicts  by \rowbufferech\%  due to the increase in memory requests sent to DRAM and (ii) P-TLB decreases row buffer conflicts by \rowbufferptlb\%  due to the reduced number of DRAM row activations for translation metadata.}
We conclude that designing more compact and efficient translation structures \konrevb{(and thus ideally approaching a perfect TLB)} can lead to a significant reduction in memory hierarchy interference.

\head{Restrictive Hash-based Mapping} 
Prior works~\cite{nearmemoryPact17,smith,mosaicpagesASPLOS2023} explore the possibility of restricting the virtual-to-physical mapping flexibility (\konreva{e.g.},
\konrevb{by computing} the physical address using a hash function applied to the virtual address)  to reduce the size of 
the data structures that store translation metadata and \konrevb{thus} reduce the address translation overhead. Restricting the virtual-to-physical mapping drastically reduces the size 
of the translation data structures, and accordingly lowers the latency of retrieving the virtual-to-physical mapping. For example, as shown in~\cite{nearmemoryPact17},
determining the physical location of a virtual page based on a specific set of bits of the virtual address is considerably faster than accessing the x86-64 multi-level PT. 
However, \konreva{restricting} the address mapping \emph{across the entire physical address space} in general-purpose systems {handicaps} core
VM functionalities and \konreva{can cause} severe performance overheads. 
First, \konreva{two} virtual pages from different processes might not be able to map to the same physical page, which limits data sharing. Second, 
the sole use of a restrictive mapping leads to increased \konrevb{swapping} activity as the system might not be able to freely map virtual pages to the available free physical space. 
\konreva{This can cause more pages to be stored inside the swap space of the storage device even in the presence of free physical memory space.}

\begin{figure}[h!]
    \vspace{-2mm}
    \centering
    \includegraphics[width=\columnwidth]{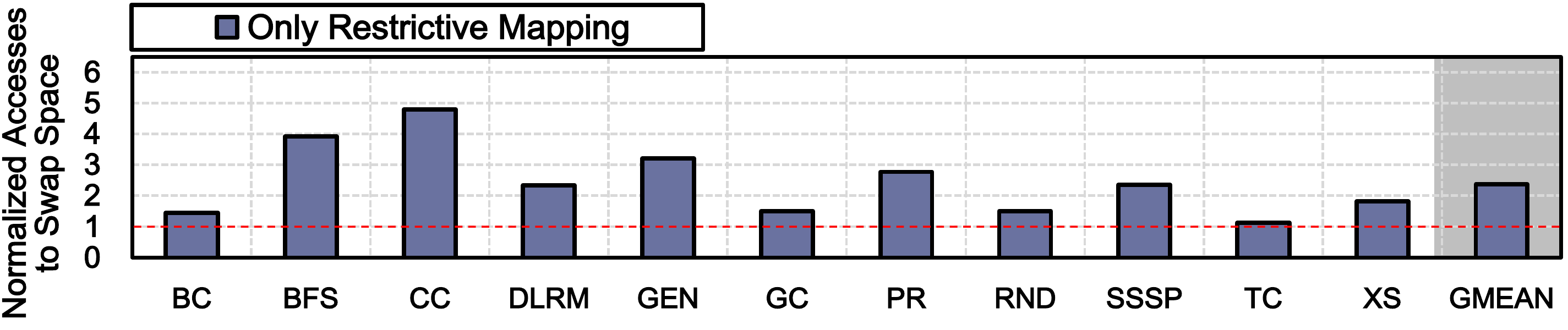}
    \vspace{-7mm}
    \caption{\konrevb{Accesses to the swap space using a system that employs a restrictive mapping across the entire memory space normalized to the baseline system.}}
    \label{fig:mot_mapping}
    \vspace{-3mm}
\end{figure}

Figure~\ref{fig:mot_mapping} shows the increase in the number of \konrevb{data accesses to the swap space} in a system that uses only the restrictive mapping across the whole memory, similar to ~\cite{nearmemoryPact17}, compared to the baseline system.  
\konrevb{We observe that} employing a restrictive address mapping \konrevb{in the entire memory space} causes a significant increase in swap space accesses, $2.2\times$ on average, since \konrevb{a large number} 
of virtual pages cannot be mapped inside physical memory and need to be stored \konrevb{into} and fetched from the swap space.
Fetching data from the swap space is orders of magnitude slower than fetching data from DRAM, which leads to significant performance overheads~\cite{flashVM2010}.

We conclude that \konrevb{neither} the sole use of a restrictive address mapping \konrevb{nor} the sole use of a flexible address mapping across the entire physical 
address space is an \konrevb{effective} solution for reducing address translation overheads while maintaining the core benefits and functionalities of VM.

\section{Utopia: Overview }
\label{sec:overview}

\subsection{Key Idea \& Design Overview}

In this work, we propose \emph{Utopia}, a new hybrid virtual-to-physical address mapping scheme that enables \emph{both} flexible and  restrictive hash-based virtual-to-physical address mapping schemes to harmoniously \emph{co-exist} in the system.
%in order to mitigate the overheads associated with address translation, while maintaining the benefits of the flexible address mapping.
\textbf{The key idea} of Utopia is to manage physical memory using two types of physical memory segments: \emph{restrictive} 
and \emph{flexible}.
%in a hybrid manner
A restrictive segment (called \emph{\utopiaseg}) enforces a restrictive, hash-based address mapping 
scheme, \konreva{thereby} enabling fast and efficient address translation through \konrevb{the use of} compact and efficient address translation structures.
A flexible segment (called \emph{\flexseg}), employs the conventional address mapping scheme and provides full virtual-to-physical address mapping flexibility. 
Utopia \konrevd{identifies and maps} costly-to-translate addresses to \utopiasegs, thereby enabling fast address translation with low translation-induced interference in the memory hierarchy 
whenever flexible address mapping is not necessary (e.g., \konreva{when optimizing for fast address translation}).
At the same time, Utopia retains the ability to use flexible address mapping to (i) support conventional VM features such as data sharing and 
(ii) avoid accesses \konreva{to the swap space} when data \konreva{does} not fit inside a restrictive segment.

\begin{figure}[h!]
    \vspace{-2mm}
    \centering
    \includegraphics[width=\columnwidth]{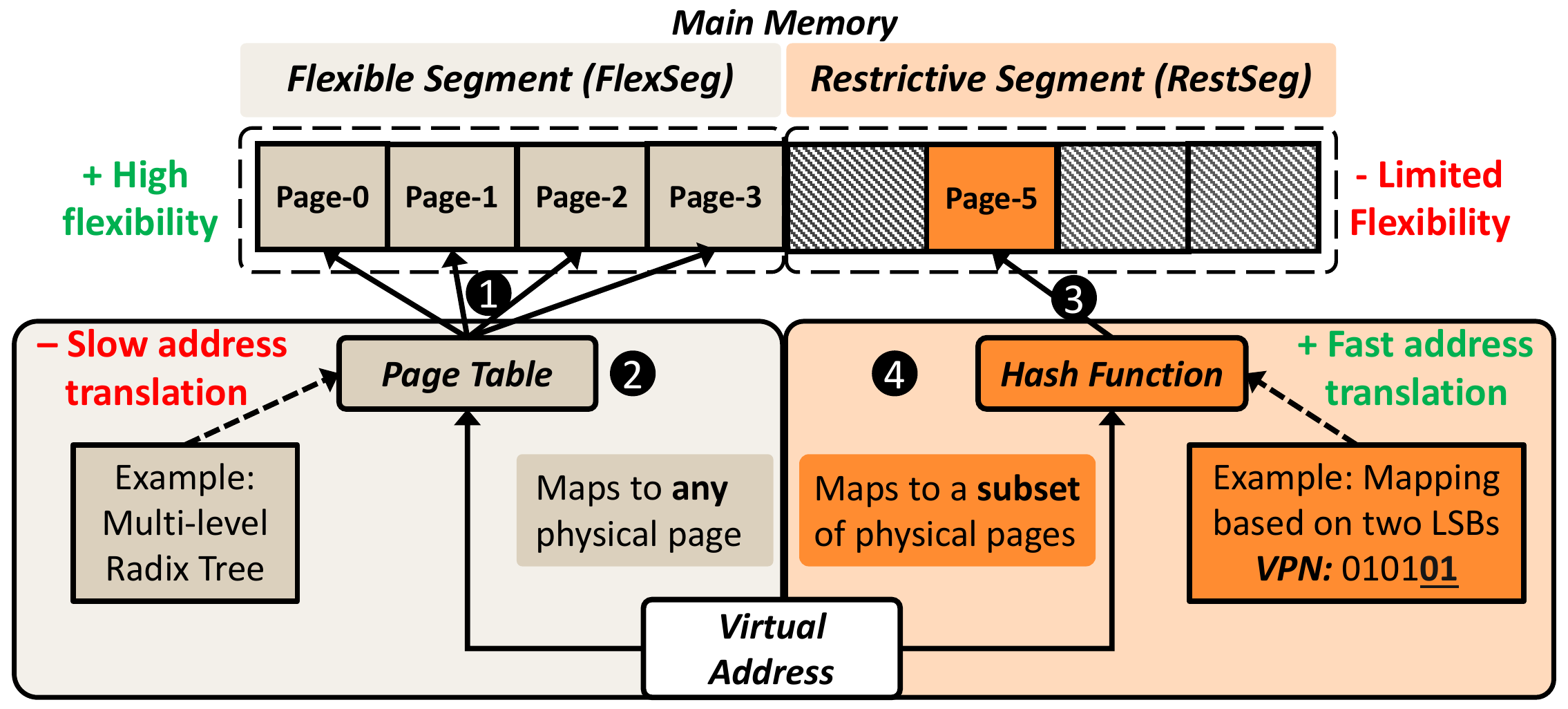}
    \vspace{-6mm}
    \caption{\konrevb{Simplified Utopia example: \flexseg vs \utopiaseg}}
    \label{fig:utopia_overview}
    \vspace{-2mm}

\end{figure}

Figure \ref{fig:utopia_overview} shows a \konrevb{simplified} example of Utopia, in which a \flexseg and a \utopiaseg co-exist in the main memory. A virtual page can be mapped to any of the physical 
pages \konrevb{(Pages 0-3)} in the \flexseg~\circled{1}, but address translation incurs high latency due to the costly memory accesses to the conventional page table~\circled{2}. 
\konrevb{In contrast}, a virtual page can be mapped to only \konrevb{one} single physical page \konrevb{(Page 5)} inside the \utopiaseg~\circled{3}, \konrevd{whose physical page number is} calculated \konrevb{using} a hash function on the virtual address (e.g., based on the LSBs of the virtual page number). 
\konrevb{Thus, the \utopiaseg} results in faster address translation but lower flexibility~\circled{4}, compared to \flexseg. In Utopia, each virtual page can reside in \emph{at most} one type of segment (i.e., either in a \flexseg or in a \utopiaseg \konrevb{but not both}). 

To perform address translation in Utopia, after an L1 TLB miss the MMU accesses in parallel (i) the translation structures of each RestSeg, i.e., an operation we call RestSegWalk (RSW), to 
discover if the data is stored in a \utopiaseg and (ii) the L2 TLB. If the data is stored neither in a \utopiaseg nor in the L2 TLB, the system initiates a conventional PTW to discover the physical location of 
the data in the \flexseg.

\subsection{Design Challenges}

To enable \konrevb{the implementation} \konrevd{of} Utopia in an efficient manner, we need to address three key challenges: (i) \konrevb{how to decide which data should be placed in a \utopiaseg}, (ii) \konrevb{create and maintain}
\utopiasegs and \flexsegs, and (iii) \konrevb{how to integrate} \papername{} in the conventional address translation pipeline.

\textbf{Challenge 1.} To address the first challenge, we design Utopia to place pages that experience high address translation latencies inside a \utopiaseg. 
To achieve that, Utopia uses two application-transparent techniques that track \emph{costly-to-translate} pages and allocate them into a \utopiaseg: (i) 
a technique that monitors the PTW cost and \konrevb{PTW} frequency of each page and \konrevb{decides if a page is costly-to-translate based on these two metrics}, 
and (ii) a page fault-based technique that directly allocates \konrevb{costly-to-translate} pages in a \utopiaseg after a page fault.

 \textbf{Challenge 2.} To address the second challenge, we extend the OS to support the (i) creation and maintenance of \utopiasegs and \flexsegs, 
 (ii) the allocation of pages inside a \utopiaseg, and  \konrevb{(iii)} the migration of data between \utopiasegs and \flexsegs. 
 In our implementation of \papername, the OS creates \utopiasegs during boot time to avoid the overhead of searching for (or creating) contiguous memory regions during runtime.

\textbf{Challenge 3.} To address the third challenge, we enhance the MMU with lightweight architectural support to integrate Utopia in the address translation pipeline. 
The MMU is extended in three ways: (i) we incorporate new hardware circuitry to enable access to the translation \konrevb{metadata} of \utopiasegs, (ii) we add two 2KB caches 
to provide fast access to the recently-used translation metadata of \utopiasegs and (iii) we parallelize the access 
to the translation \konrevb{metadata} of \utopiasegs  with the L2 TLB access to reduce  address translation latency.

\section{Utopia: Detailed Design}
\label{sec:detailed}
\konrevb{We describe} in detail 
(i) the key properties of a \utopiaseg, 
(ii) how to \konrevb{perform address} translation for pages that reside in a \utopiaseg, 
(iii) how to resolve \konrevb{address translation} in the presence of hybrid address mapping, 
(iv) how Utopia \konrevb{decides} which data should be placed into a \utopiaseg \textit{(Challenge 1)}, 
(v) the OS extensions to enable Utopia \textit{(Challenge 2)}, and 
(vi) the architectural modifications in the MMU to efficiently support Utopia \textit{(Challenge 3)}.

%In this section, we discuss the key properties of a \utopiaseg, address translation resolution for \utopiaseg pages, and the lightweight MMU modifications to efficiently support Utopia.
%We also address the challenges of data placement detection and OS extensions for Utopia.

\subsection{Segment with Restrictive Address Mapping}
\label{sec:structure}
We design \utopiaseg as a physical memory segment that enforces a \emph{set-associative address mapping}. 
A virtual page can map only to a specific set of physical pages in the \utopiaseg, in a similar way that \konrevc{set-associative} 
caches store \konrevc{sets of} cache blocks \konrevc{at a particular index value}.\footnote{\konrevc{i.e., a memory block can map to a specific set of cache blocks.}}
The set-associative address mapping accelerates address translation for virtual pages mapped in a \utopiaseg since discovering a 
virtual-to-physical mapping only requires calculating the set index using a hash operation \konrevc{on the virtual address} followed by a tag matching operation. 
In this section, we discuss the key properties of a \utopiaseg and the translation structures that enable efficient address translation for pages that reside in a \utopiaseg.

\subsubsection{\noindent Key Properties of \utopiaseg\newline}
\utopiaseg's key properties enable (i) adaptability to diverse workloads and system configurations, 
 (ii) backward compatibility with existing OS primitives, and (iii) efficient address translation for multiple processes at the same time.

\begin{figure}[h!]
    \vspace{-2mm}
    \centering
    \includegraphics[width=\columnwidth]{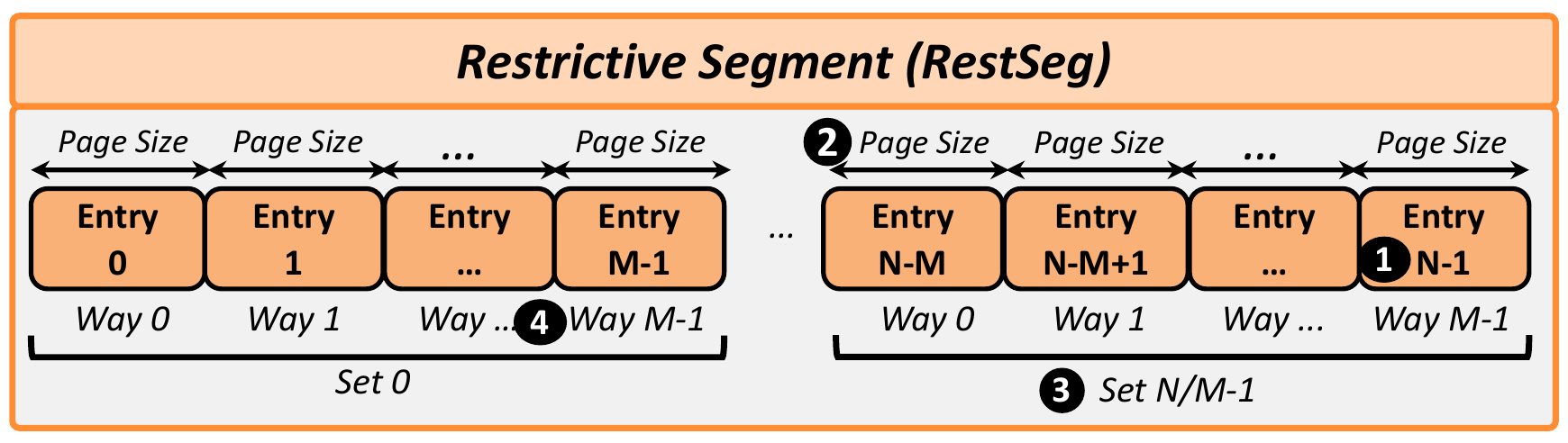}
    \vspace{-7mm}
    \caption{Structural properties of a \utopiaseg.}
    \label{fig:structural}
    \vspace{-2mm}
\end{figure}

\head{Structural Properties of \utopiaseg} \konrevd{Figure~\ref{fig:structural} shows the structure of a \utopiaseg.
\utopiaseg is a contiguous physical memory segment of associativity $M$
that contains $N$ physical pages~\circled{1} of equal size~\circled{2} (e.g., 4KB) organized in $N/M$ sets~\circled{3} of $M$ ways~\circled{4}.
Each virtual page can map to any of the $M$ ways of the set that it corresponds to.
\konrevs{Different} \utopiasegs can be configured at their creation with different size, page size and associativity (as we discuss in \S\ref{sec:os-support}, a \utopiaseg is created 
during boot time) to adapt to a diverse set of workloads and system configurations.}

\head{Multiple \utopiasegs in the System}
In our implementation, Utopia uses two RestSegs, one that stores 4KB and one that stores 2MB pages to retain backward compatibility with 
existing large page mechanisms~\cite{corbet2011,arcangeli2010}.
However, it is possible to employ more than two RestSegs in a single system to satisfy the needs of different workloads 
(e.g., to support three different page sizes~\cite{tridentMICRO2021} or \konrevd{relieve a fully-allocated RestSeg from memory capacity pressure).}

\head{Sharing \utopiaseg Across Processes} A \utopiaseg can store pages from different processes since the translation structures of a \utopiaseg are stored per process.
Sharing a \utopiaseg across multiple processes is useful in scenarios where multiple processes benefit from the fast address translation of a \utopiaseg.
At the same time, sharing a \utopiaseg across multiple processes can lead to efficient memory utilization 
when the processes have different memory footprints and the \utopiaseg is not fully utilized by a single process.

\subsubsection{Translation Structures of \utopiaseg\newline} To locate a page in a \utopiaseg, we introduce two new translation structures:
the Tag Array (TAR) and the Set filter (SF). Figure~\ref{fig:structural} shows the TAR and SF for an example 4-entry 2-way associative \utopiaseg.

\begin{figure}[h!]
    \vspace{-2mm}
    \centering
    \includegraphics[width=\columnwidth]{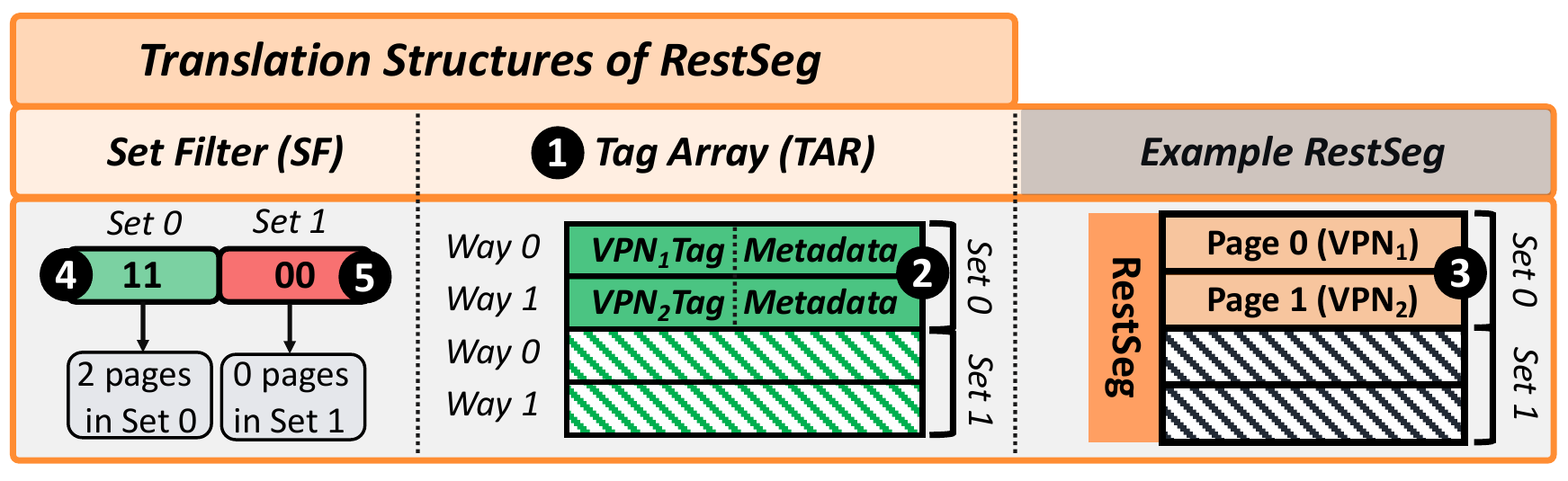}
    \vspace{-7mm}
    \caption{Translation structures of a \utopiaseg.}
    \label{fig:sftar-restseg}
    \vspace{-2mm}
\end{figure}

\head{Tag Array (TAR)} TAR stores the tags of every way of every set of the \utopiaseg~\circled{1}. 
Each tag consists of the virtual page tag and 10 extra bits for various metadata (e.g., access permissions)~\circled{2}.
In the example of Fig.~\ref{fig:structural}, TAR stores the tags and metadata of the ways 0 and 1 of set 0 since 
only these two ways of the RestSeg are occupied~\circled{3}.
Each virtual page tag requires 48-$log_2(4KB)$-$log_2(2)$ = 48-12-1 = 35 bits and the total size of the TAR is (4 $\times$ (35+10))/8 = 180 bits. \footnote{\sloppy
The general formula to compute the size of the TAR is the following: 
\begin{align*}
    \left(\frac{\text{RestSegSize}}{\text{PageSize}}\right) \times \left(48 - \log_2(\text{PageSize})
    - \log_2\left(\frac{\text{RestSegSize}}{\text{PageSize}\times\text{Associativity}}\right)+10\right)
\end{align*}}

\head{Set Filter (SF)} \konrevd{SF is used to quickly discover if a set of the \utopiaseg is empty (i.e., all ways are empty) or not. 
SF stores an array of $\#sets$ counters of length $log_2(assoc)+1$ that keep track of the cardinality of every set of the \utopiaseg.
Each counter gets incremented/decremented when a new page of a process is added/removed from the corresponding set.
In the example of Fig.~\ref{fig:structural}, SF stores two 2-bit counters, one for set 0 and one for set 1~\circled{4}.
Thus, the total size of the SF is $(16KB/4KB/2)\times$ ($log_{2}(2)+1) = $ 4 bits. \footnote{\sloppy
The general formula to compute the size of the SF is:
\begin{align*}
\left(\frac{\text{RestSegSize}}{\text{PageSize}\times\text{Associativity}}\right) \times \left(\log_2(\text{Associativity})+1\right).
\end{align*}}
The counter of set 0 is equal to 11 since two pages are stored in set 0~\circled{5}. 
The counter of set 1 is equal to 00 since no pages are stored in set 1~\circled{5}.}

\begin{figure}[h]

    \centering
    \includegraphics[width=\linewidth]{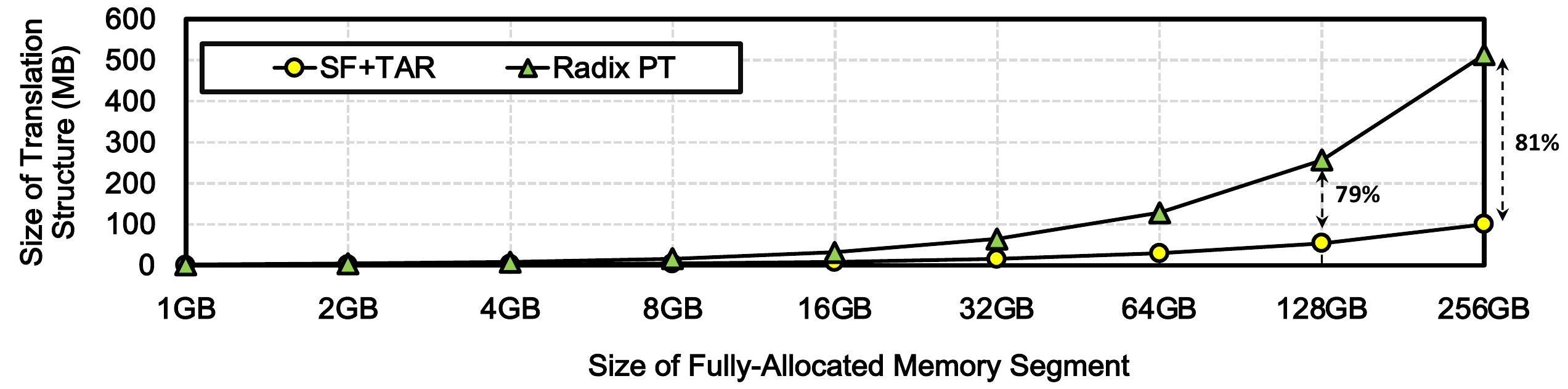}
    \vspace{-7mm}
    \caption{Memory consumption of TAR/SF vs. conventional multi-level radix-based page table.}
    \label{fig:tarsfvspt}
    \vspace{-2mm}
\end{figure}

\head{Scalability of TAR/SF} Figure~\ref{fig:tarsfvspt} shows how the size of TAR/SF scales compared to the radix-based page table, \konrevd{across fully-allocated physical memory segments
(i.e., all physical pages are occupied)} of increasing sizes.
We observe that for the largest allocated memory size (256GB) TAR and SF consume 81\% less memory than the radix-based page table.
We conclude that Utopia's \konrevd{new} translation structures (TAR and SF) scale efficiently as the size of allocated memory increases.

\head{Storing TAR/SF} TAR and SF are stored in kernel memory per process per \utopiaseg to provide process isolation guarantees (i.e., a process cannot access the TAR/SF of another process).
\konrevd{The OS employs \emph{a global TAR} (stored in kernel memory) to maintain a global view of pages that reside in a \utopiaseg across all processes.\footnote{TAR/SF and the conventional page table store completely-disjoint virtual-to-physical mappings to avoid complex coherence operations between the translation structures.}
The OS uses the global TAR to discover and allocate
free pages in a \utopiaseg when needed.}

\subsection{Address Translation for Data in RestSeg}

\begin{figure}[h]
    \vspace{-2mm}
    \centering
    \includegraphics[width=\linewidth]{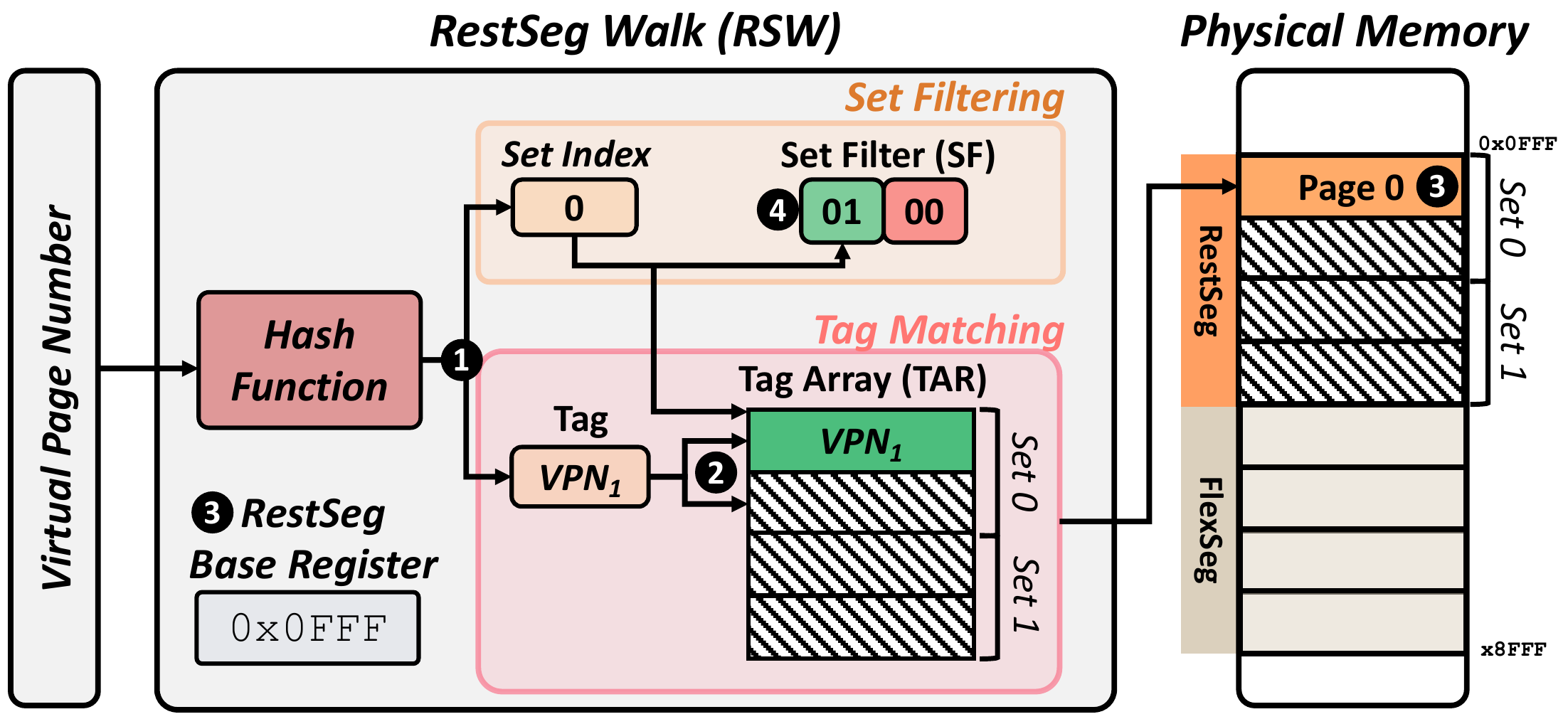}
    \vspace{-6mm}
    \caption{ RestSeg Walk: Address translation in \utopiaseg}
    \label{fig:utopia_walk}
    \vspace{-2mm}
\end{figure}

\utopiaseg uses TAR and SF to discover the physical location of a virtual page inside the physical memory space. 
We call this process \utopiaseg Walk (RSW). RSW consists of two operations: (i) tag matching and (ii) set filtering.
Figure \ref{fig:utopia_walk} shows the operations of RSW in a system that employs a 4-entry 2-way associative \utopiaseg that stores 4KB pages.

\head{Tag Matching} To perform tag matching for a virtual page, a hash function is applied to the virtual page number (VPN) 
to retrieve the set index and the virtual page tag (virtual page tag is equal \konrevd{to} $VPN_1$ and set index is equal to 0 in Fig.~\ref{fig:utopia_walk}~\circled{1}).
TAR is looked up to compare the virtual page tag with the tags of all ways of the corresponding set~\circled{2}. 
If the virtual page tag matches with the tag of way $i$, the virtual page resides in way $i$ ($tar[0]=VPN_1$ and $i=0$). 
The physical address is computed directly as $RestSegBaseRegister + set\_index \times associativity+i$ where the \konrevd{\utopiaseg Base Register} points to the 
beginning of the \utopiaseg in the physical address space ($VPN_1$ is stored in physical page 0 in physical address $0x0FFF$~\circled{3}). 
If the virtual page tag does not match with any of the tags \konrevd{(stored in any of the ways)} in the set, the virtual page does \emph{not} reside in the \utopiaseg.

\head{Set Filtering} For every tag matching operation, all the tags of the set need to be \konrevd{looked up even} if they are invalid (i.e., the set is empty).
To avoid \konrevd{looking up the TAR when sets are empty}, RSW \konrevd{looks up} the SF to quickly discover if a virtual page does not reside in the \utopiaseg. 
The SF is indexed using the set index to retrieve the counter of the corresponding set ($set\_index=0$ and $set\_filter[0]=01$~\circled{4}).
If the counter is $0$, tag matching is skipped since all ways of the set are invalid (i.e., the virtual page does not reside in the \utopiaseg).
If the counter is $>0$, the system proceeds to tag matching to identify whether the virtual page is 
kept in any way of the set (the set contains $VPN_1$). 
SF comes with two key benefits: (i) SF is smaller than TAR and hence enjoys \konrevd{better} temporal locality than TAR 
\konrevd{(ii) \konrevs{SF} enables skipping tag matching for sets that are empty, thereby avoiding expensive lookups to TAR.}

RSW provides two key benefits over the conventional four-level PTW.
First, RSW  generates only two \emph{parallel} memory accesses: \konrevd{one} for SF and \konrevd{one for} TAR. 
In contrast, a conventional PTW always requires four \emph{sequential} memory accesses.
Second, as we show in \S\ref{sec:evaluation}, TAR and SF take better advantage of hardware caching as they are smaller 
and exhibit high spatial/temporal locality compared to the PT. 
 
% Fig. \ref{fig:utopia_walk} shows an example of performing an RSW for page 0 that resides in a 2-way associative 16KB \utopiaseg. 
% First, the VPN of page 0 is passed through a hash function to retrieve the set index (0 in this case) and the virtual tag \circled{1}. 
% Then, the set filtering \circled{2} and the tag matching happens in parallel \circled{3}. 
% For page 0, tag matching is successful, and the physical address is computed directly as ($BaseRegister + set\_index \times associativity+\#way$) where the base register points to the 
% beginning of the \utopiaseg in the physical address space \circled{6}. 

\subsection{Segment with Flexible Address Mapping}

The structure of \flexsegs is similar to that of the conventional flexible segments used in modern VM designs 
(which we discussed in detail in \S\ref{sec:background}). \flexseg uses a fully-flexible virtual-to-physical address mapping: a
virtual page can map to any physical \konrevd{page}. We call the process of looking up the PT to perform address translation FlexSeg Walk (FSW) 
(same as PTW in modern VM).\footnote{We use the term FSW to clearly distinguish the process of address translation in a \flexseg from the process of address translation in a \utopiaseg.}
For each process in the system, the OS maintains a single PT that stores the virtual-to-physical mappings of all the pages of the process,
regardless of the number of \flexsegs that the process uses.\footnote{The system does not use a separate PT per \flexseg since the virtual-to-physical mappings of all \flexsegs can be stored in the same PT.}

% \begin{figure}[h]
%     \vspace{-2mm}
%     \centering
%     \includegraphics[scale=0.3]{figures/utopia_organization.pdf}
%     \vspace{-6mm}
%     \caption{ Example memory organization in Utopia}
%     \label{fig:utopia_organization}
%     \vspace{-2mm}
% \end{figure}

\subsection{Address Translation \konrevd{Flow} in Utopia}

Figure~\ref{fig:workflow} shows a high level description of how address translation is performed in a system that employs Utopia with two \utopiasegs, one that stores 4KB pages and one that stores 2MB pages, and one \flexseg.
On an L1 TLB miss, the system in parallel (i) performs two \utopiaseg walks, one for each RestSeg (i.e., one for the RestSeg that stores 4KB pages~\circled{1} and one for the RestSeg 
that stores 2MB pages~\circled{2}) and (ii) probes the L2 TLB~\circled{3}. If the physical address is found \konrevd{during} either RSWs~\circled{4} or in the L2 TLB~\circled{5},
the translation request is resolved without performing an FSW. If not, an FSW is performed~\circled{6} to discover the physical 
address~\circled{7}.

%In Section~\ref{sec:arch-support}, we discuss in detail how we design the MMU to overlap the RSW with the L2 TLB access and reduce address translation latency. 
% To translate a virtual address in a system with both \utopiasegs and \flexsegs, Utopia needs to discover if the virtual \rbc{page} is stored in a \utopiaseg or a \flexseg (a virtual page can only reside in one of the two different segments). All \flexsegs in the system use the same page table (PT) per process. Hence, discovering whether a page resides into a \flexseg is done by accessing the PT. On the other hand, discovering \rbc{whether} a virtual \rbc{page} resides in any of the $m$ \flexsegs of the system requires performing $m$ Utopia Walks. Putting it all together, Utopia translates a virtual address by performing one PTW and $m$ RSWs in parallel.

% \begin{figure*}[h]
%     \centering
%     \includegraphics[scale=0.36]{figures/mmu_utopia.pdf}
%     \vspace{-4mm}
%     \caption{(a) Address translation pipeline in Utopia. (b) MMU extensions to support Utopia.}
%     \label{fig:workflow}
%     \vspace{-4mm}
% \end{figure*}

\begin{figure}[h]
    \vspace{-2mm}
    \centering
    \includegraphics[width=\linewidth]{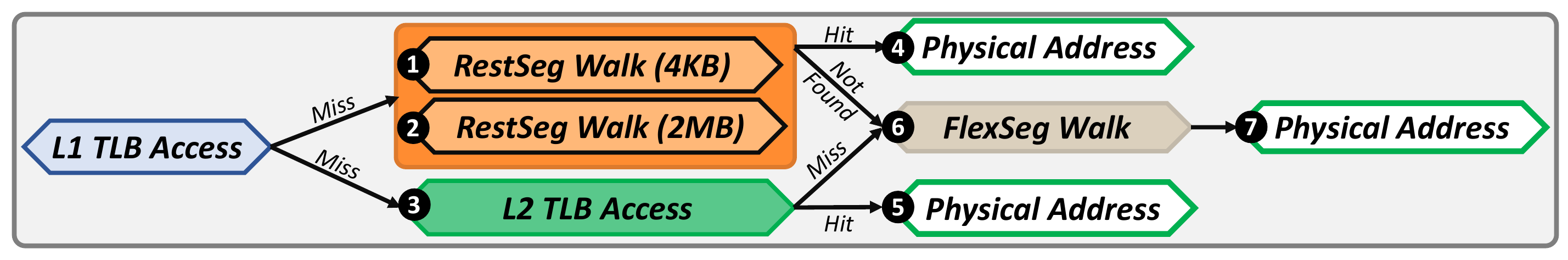}
    \vspace{-7mm}
    \caption{\konrevd{Address translation flow in Utopia.}}
    \label{fig:workflow}
    \vspace{-3mm}
\end{figure}

\subsection{Data Placement in a \utopiaseg }
\label{sec:heuristics}

We design Utopia to place pages that experience high address translation latencies inside a \utopiaseg. 
To achieve that, Utopia uses two application-transparent techniques to allocate pages into a \utopiaseg: 
(ii) a Page-Fault-based technique that directly allocates pages inside a \utopiaseg and 
(i) a PTW-Tracking-based technique that migrates pages to a \utopiaseg based on the PTW cost and frequency of each page.  

\head{Page-Fault-based Page Allocation in the RestSeg}
\konrevd{This technique directly allocates a page in a 
\utopiaseg when \konrevd{servicing} a page-fault (PF) \konrevd{for that page}. The key insight is to treat all pages as \konrevd{costly-to-translate} from the get go
and allocate them in a \utopiaseg to avoid the cost of a future PTW. 
\konrevs{Utopia allocates a page in a \flexseg in two cases: (i) when the corresponding page gets evicted from a \utopiaseg or
(ii) when there is not \konrevf{enough free} memory space in any \utopiaseg.}}

\head{PTW-Tracking-based Page Migration into the RestSeg} The key idea is to monitor the 
PTW frequency and PTW cost of each virtual page stored in the FlexSegs and migrate pages that experience high PTW frequency and cost to a \utopiaseg.
To do so, two additional counters stored in the unused bits of the PTE (9 bits in x86-64~\cite{intelx86manual}).
Whenever a translation request misses in the L2 TLB, the PTW frequency counter of the corresponding PTE is incremented and the
PTW cost counter is increased by the number of DRAM accesses performed by the PTW. 
When both the PTW frequency counter and the cost counter exceed pre-determined threshold \konrevd{values} (which can be configured using programmable registers), 
the page is migrated from the FlexSeg to a \utopiaseg. After migrating the page to the \utopiaseg, 
the corresponding entries are erased from the PT of the FlexSeg and the counters are reset.

\subsection{Operating System Support for Utopia}
\label{sec:os-support}

\head{Creation of \utopiaseg}
The OS creates the \utopiasegs during boot time. \konrevd{Doing so avoids the runtime overheads}
of compacting memory during runtime in order to create contiguous physical memory segments.

\head{Data Allocation in a \utopiaseg}
\konrevd{The OS directly allocates a page in a \utopiaseg when servicing a page-fault (PF) for that page (as we describe in \S\ref{sec:heuristics}).
During the page fault, a hardware interrupt hands the control to the OS. 
The OS computes the set index of the page in the RestSeg by applying \konrevs{a} hash function to the virtual page number (VPN).
Using the set index, the OS accesses the global TAR to search for a free way in the set. If the set has a free way, 
the OS places the page in the set and updates the TAR and SF of the process.}

\head{Eviction of a Page from \utopiaseg to FlegSeg}
During a page allocation in a \utopiaseg, \konrevs{if} the corresponding set has no free ways, the OS handles the conflict and evicts a page from the set. 
The OS employs a replacement policy (we use SRRIP~\cite{srrip} in our evaluation) to decide which page to evict from the set. 
The OS triggers a page migration to move the evicted page from the \utopiaseg to a \flexseg.
When the migration is complete, the OS updates the translation structures of the \utopiaseg and the global TAR.

\head{Migration of a Page from FlexSeg to \utopiaseg}
When the PTW-Tracking migration policy discovers a costly-to-translate page (as we describe in \S\ref{sec:heuristics}), 
the MMU sends an asynchronous interrupt to the OS, so that the OS migrates the page into the RestSeg without stopping program execution. 
If there is free space in the corresponding set of the RestSeg, a single migration is performed, from the FlexSeg to the RestSeg. 
If there is no free space in the corresponding set of the RestSeg, the OS performs (i) the migration of the costly-to-translate page from the FlexSeg to the RestSeg and
(ii) the migration of the evicted page from the RestSeg to the FlexSeg.

\begin{figure}[h]
    \centering
    \includegraphics[width=\linewidth]{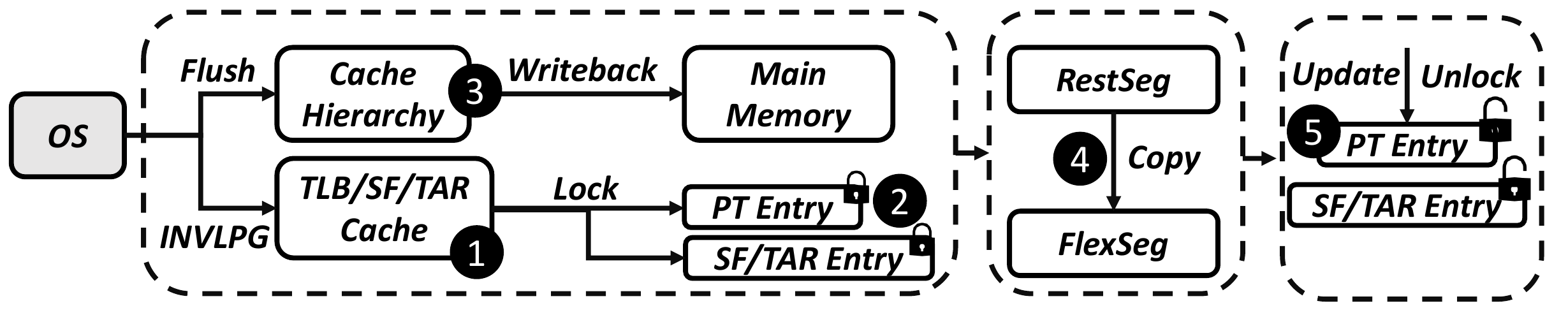}
    \vspace{-7mm}
    \caption{Steps followed during a page migration.}
    \vspace{-3mm}
    \label{fig:mig_detail}
\end{figure}

\head{Performing a Page Migration}
The OS takes four steps to migrate a page to/from a RestSeg to/from a FlexSeg.
Figure~\ref{fig:mig_detail} shows the migration process of a page from a \utopiaseg to a \flexseg and vice versa.
First, the OS performs a TLB shootdown (and a TAR/SF cache shootdown, as described in \S\ref{sec:arch-support})~\circled{1} to maintain coherence and 
locks the corresponding entries of the translation tables~\circled{2} to make sure the program cannot access the translation tables while the migration is happening.  
Second, the OS flushes all the dirty cache lines of the migrated page from the cache hierarchy~\circled{3} to ensure that the data in the main memory is not stale.
\konrevs{Third, the OS copies the page to the destination memory region using the Direct Memory Access (DMA) engine to avoid stalling the CPU during the migration process~\circled{4}}.\footnote{
\konrevf{Such page} copy \konrevw{operations} can be accelerated using in-DRAM \konrevf{data copy mechanisms as proposed and evaluated by 
~\cite{seshadri2013,computeDRAMmicro2019,pidram,chang2016low,nom}. We do not assume the existence of such mechanisms in the system
and leave the evaluation of in-DRAM~\konrevf{data copy} in Utopia to future work.
}}
Fourth, the OS updates all the translation tables, PT, TAR and SF and "unlocks" them so that the application
can fetch the corresponding cache lines from the cache hierarchy~\circled{5}. While the migration happens, the cache lines of the migrated page cannot be
accessed by any running application.

\subsection{Architectural Support in MMU for Utopia }
\label{sec:arch-support}

The MMU is extended with a hardware-based \utopiaseg walker to accelerate address translation for data in \utopiasegs.
RestSeg walker consists of two components:
(i) a new hardware FSM that can access the TAR and SF of the \utopiasegs and 
(ii) two (2$\times$2KB) SRAM caches, \konrevd{the TAR and SF cache}, that store recently-accessed TAR and SF entries. 
The \konrevd{address} translation flow of the new MMU is shown in Figure~\ref{fig:workflow}.
We describe how the MMU performs address translation in every possible scenario: 
(i) the address mapping is cached in the TLB hierarchy (\textit{TLB hit}), 
(ii) data is stored in a \utopiaseg and the physical address is determined by the \utopiaseg walker (\textit{RestSeg Walk}) and
(iii) data is stored in a \flexseg, the address mapping is \emph{not} cached in the TLB hierarchy and the physical address is determined by the FlexSeg walker (\textit{FlexSeg walk}).

\begin{figure}[h]
    \vspace{-2mm}
    \centering
    \includegraphics[width=\linewidth]{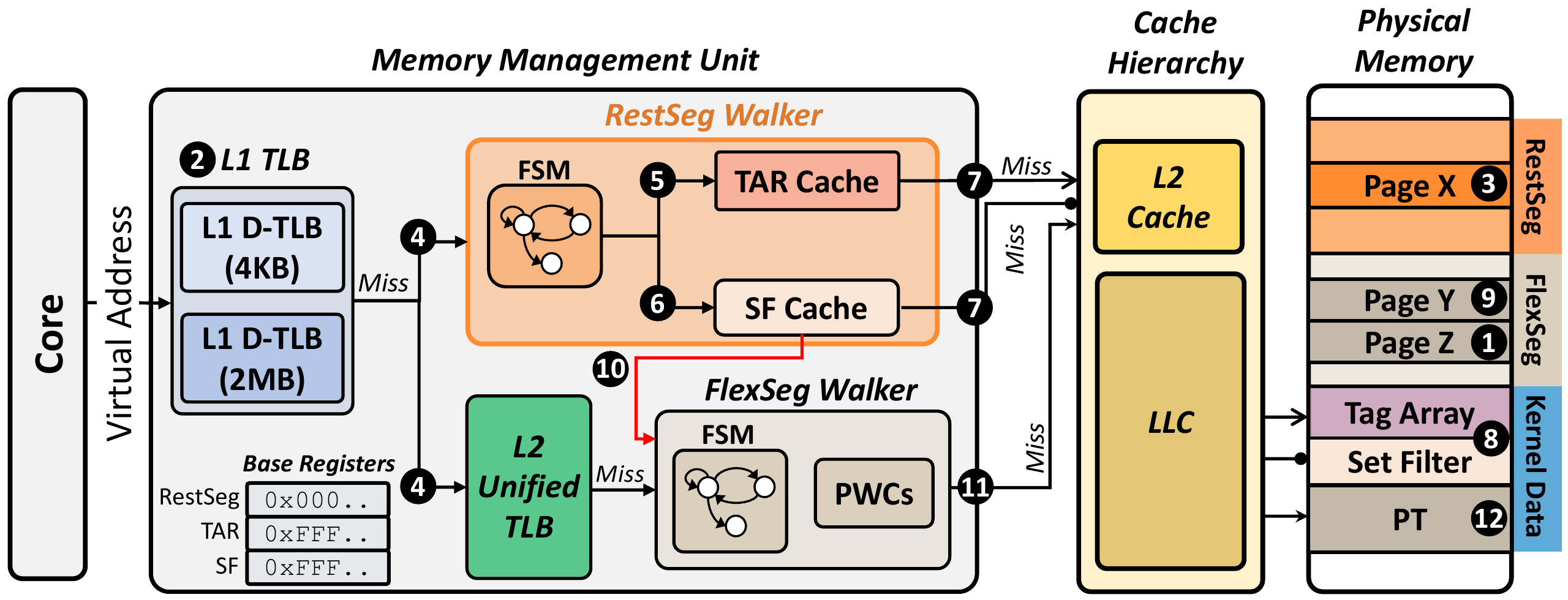}
    \vspace{-6mm}
    \caption{MMU extensions and operations to support Utopia.}
    \label{fig:workflow}
    \vspace{-2mm}
\end{figure}

\head{TLB Hit} The core accesses page Z which is stored in a FlexSeg~\circled{1}. The MMU looks up in parallel the L1 TLBs (4KB and 2MB) with the VPN of Z. 
\konrevd{If the virtual-to-physical mapping} is found in the L1 TLB~\circled{2} the physical page number (PPN) is sent to the core 
(same happens if the mapping is found in the L2 TLB).

\head{RestSeg Walk} The core accesses page X which is stored in a \utopiaseg~\circled{3}.
The MMU misses \konrevd{in} the L1 TLB and in parallel~\circled{4} (i) looks up the L2 TLB and (ii) triggers an RSW. 
The RestSeg walker accesses in parallel the SF cache~\circled{5} and the TAR cache~\circled{6}. \konrevs{The base 
addresses of TAR and SF are stored in two additional registers inside the MMU.}
\konrevd{On a TAR/SF cache hit, the corresponding TAR/SF entries are looked up without accessing the memory hierarchy.}
On a TAR/SF cache miss~\circled{7}, the corresponding TAR/SF entries are fetched into the TAR/SF cache from the cache hierarchy or the main memory~\circled{8}. 
In this scenario, page X resides in \utopiaseg~\circled{3}. 
Thus, RSW delivers the virtual-to-physical mapping to the core and the MMU aborts the L2 TLB access and the FSW.

\head{FlexSeg Walk} The core accesses page Y~\circled{9} which is stored in a \flexseg. 
The address mapping is not found in the L1 and L2 TLBs and an RSW is initiated in parallel with the L2 TLB~\circled{10}. 
The FSW is stalled until the RSW confirms whether or not the page resides in a \utopiaseg~\circled{10}. 
\konrevd{If not,} the RestSeg walker informs the FlexSeg walker that the page does not reside in a \utopiaseg 
and the FSW \konrevd{is initiated to provide} the virtual-to-physical mapping. 
\konrevd{The FlexSeg walker accesses the PWCs to look for the intermediate levels of the PT.
On PWC misses~\circled{11}, the FlexSeg walker fetches the PT from main memory~\circled{12}.}

\section{System Integration}

\subsection{\konrevf{Context Switches} in Utopia}
The MMU accesses the translation structures of a \utopiaseg (TAR and SF) of a process using specialized registers that store the base 
address of each data structure (one for TAR and one for SF), similarly to how the CR3 register works \konrevf{in the x86-64 ISA}. 
After a context switch, the TAR and SF registers are \konrevf{reloaded} by the OS to point to the translation structures of the new process.
The contents of the TAR and SF caches do \emph{not} need to be flushed during a context switch since both caches
operate using physical addresses.

In conventional systems, whenever a virtual-to-physical mapping gets modified \konrevf{(e.g., due to a page migration or a page deallocation)}, 
all the affected TLB entries of  all the running processes are invalidated to maintain TLBs coherent. 
When Utopia triggers a \konrevf{page} migration to/from a RestSeg from/to a FlexSeg, the OS gets invoked and sends an INVLPG instruction~\cite{inteltlb} to the MMU. 
The corresponding TAR/SF cache entries are invalidated to maintain TAR/SF caches coherent, \konrevd{in the} same way as TLB entries are invalidated in conventional systems.
To invalidate the TAR/SF entries \konrevf{that correspond to the modified virtual-to-physical mapping}, 
MMU computes the indices to the TAR/SF data structures by applying a hash function on the virtual address \konrevf{of the modified page.}
The TAR/SF caches are probed using the calculated indices and the corresponding entries get invalidated.

\subsection{Area \& Power Overhead}
Utopia extends the MMU with two 2KB SRAM structures, \konrevf{the TAR cache and SF cache (\S\ref{sec:arch-support})}. We measure the area and power 
overhead of extending the MMU to support Utopia using the $45$-nm library of McPAT~\cite{mcpat} and compare it against an Intel Raptor Lake CPU~\cite{raptor_lake}. 
Utopia incurs an area and power overhead of \area\% and \power\% \konrevf{per core}, respectively.

% \begin{figure*}[ht]
%     \centering
%     \includegraphics[width=6.8in]{figures/4plot.pdf}
%     \vspace{-3mm}
%     \caption{(a) Performance speedup of Utopia-PF-PTE and No-Translation compared to the Baseline system (b) Fraction of LLC blocks that contain translation entries in Baseline and Utopia-PF-PTE (c) LLC MPKI of data requests in Baseline and Utopia-PF-PTE (d) Number of migrations caused by Utopia}
%     \label{fig:single_core_four_plot}
% \end{figure*}

\section{Evaluation Methodology}
\label{sec:methodology}

We evaluate Utopia using the Sniper Simulator \cite{sniper}. \konrevd{This simulator is freely available at \textcolor{blue}{\url{https://github.com/CMU-SAFARI/Utopia}}.}
We extend Sniper to accurately model: 
(i) TLBs that support multiple page sizes,
(ii) the conventional radix page table walk, 
(iii) page walk caches,
(ii) memory management (e.g., allocation using the buddy allocator~\cite{buddy}), 
(iii) page migration latency, and 
(iv) the functionality and timing of all the evaluated systems. 
Table \ref{tab:simconfig} shows the simulation configuration of (i) the baseline system and (ii) all evaluated systems.

% \konkanelloreva{The essential OS operation that needs to be modeled to properly evaluate Utopia is the migration of pages between RestSeg/FlexSeg. 
% We extend Sniper to model the page migration and its effects on performance as follows. First, the model determine the latency of a page migration 
% due to DRAM read/write operations (~1000 cycles, details in Table~\ref{table:sim_params}). Our model assumes all cache lines in the migrated page 
% are unavailable during the course of the page migration. Second, we invalidate the corresponding translation entries in the L1-L2-TLB, SF/TAR caches. 
% Third, it flushes the data of the migrated page from caches.  }

% \usepackage{color}
% \usepackage{tabularray}
% \usepackage{color}
% \usepackage{tabularray}
% \usepackage{color}
% \usepackage{tabularray}
\definecolor{SoftPeach}{rgb}{0.937,0.901,0.901}
\begin{table}[h!]
\centering
\scriptsize
\caption{Simulation Configuration and Simulated Systems}
\vspace{-2mm}
\label{tab:simconfig}
\begin{tblr}{
  width = \linewidth,
  colspec = {Q[220]Q[619]},
  row{1} = {SoftPeach,c},
  row{15} = {SoftPeach,c},
  cell{1}{1} = {c=2}{0.94\linewidth},
  cell{3}{1} = {r=4}{},
  cell{7}{1} = {r=2}{},
  cell{9}{1} = {r=2}{},
  cell{12}{1} = {r=2}{},
  cell{15}{1} = {c=2}{0.94\linewidth},
  cell{16}{1} = {r=2}{},
  cell{18}{1} = {r=2}{},
  cell{21}{1} = {r=2}{},
  vlines,
  hline{1-3,12,14-16,21,24} = {-}{},
  hline{4-11,13,17-20,22-23} = {-}{},
}
\textbf{Baseline System} & \\
\textbf{Core} & 4-way OoO x86 2.6GHz core\\
\textbf{MMU} & L1 I-TLB: 128-entry, 8-way assoc, 1-cycle latency\\
 & {L1 D-TLB (4KB): 64-entry, 4-way assoc, 1-cycle latency \\ L1 D-TLB (2MB): 32-entry, 4-way assoc, 1-cycle latency}\\
 & L2 TLB: 1536-entry, 12-way assoc, 12-cycle latency\\
 & 3 Split Page Walk Caches: 32-entry, 4-way assoc,  2-cycle latency\\
\textbf{L1 Cache} & {L1 I-Cache: 32~KB, 8-way assoc, 4-cycle access latency \\ L1 D-Cache: 32~KB, 8-way assoc, 4-cycle access latency}\\
 & LRU replacement policy;~IP-stride prefetcher~\cite{stride}\\
\textbf{L2 Cache} & 2 MB, 16-way assoc, 16-cycle latency\\
 & SRRIP replacement policy~\cite{srrip}; Stream prefetcher~\cite{streamer}\\
\textbf{L3 Cache} & 2~MB/core, 16-way assoc, 35-cycle latency\\
\textbf{Main Memory} & 32~GB, DDR4-3200  $t_{RCD}$=12.5ns, $t_{RP}$=2.5ns \newline $t_{CL}$=12.5 ns , $t_{RTP}$= 7.5ns \\
& Migration Latency: 2 DRAM Full row reads/writes \newline 2$\times$($t_{RCD}$+ $t_{CL}$/$t_{CWL}$$\times$(64$\times$64B) + $t_{RTP}$ + $t_{RP}$)$\approx$1000 cycles \\
\textbf{Transparent Huge \newline Pages (THP) ~\cite{arcangeli2010,corbet2011} } & {Debian 9 4.14.2. 10-node cluster \\ Memory per node: 256GB-1TB}\\
\textbf{\textbf{Evaluated Systems}} & \\
\textbf{POM-TLB~\cite{pomtlbISCA2017}} & 64K-entry L3 software-managed TLB, 16-way assoc\\
 & SRRIP replacement policy~\cite{srrip}\\
\textbf{Elastic Cuckoo Hash Table (ECH)~\cite{elastic-cuckoo-asplos20}} & {8192-entries/way, 4-way, Scaling: 10, Occupancy: 0.6, \\Hash function: CITY~\cite{cityhash} 2-cycle lat.,}\\
 & 2$\times$16-entry Cuckoo Walk Caches, 2-cycle latency\\
\textbf{RMM~\cite{karakostas2015}} & 32-entry Range Lookaside Buffer, Eager paging allocator\\
\textbf{Utopia} & 2 x 512MB \utopiasegs: 1$\times$4KB pages and 1$\times$2MB pages \newline RestSegs: 16-way, SRRIP repl. policy~\cite{srrip} \newline  1x FlexSeg with x86-64 4-level radix page table \\
 & TAR Cache: 2KB, 2-cycle latency, SF Cache: 2KB, 2-cycle latency \newline Hash function: modulo hash\\
\textbf{Perfect TLB (P-TLB)} & Translation requests always hit in the L1 TLB\\
\end{tblr}
\end{table}

\textbf{Workloads.} Table~\ref{tab:workloads} shows all the benchmarks we use to evaluate Utopia and the systems we compare Utopia to. 
We select applications with high L2 TLB MPKI ($>5$), which are also used in previous works~\cite{elastic-cuckoo-asplos20,compendiaISMM2021,midgard,flataAsplos2022}. 
We evaluate our design using seven workloads from the GraphBig \cite{Lifeng2015} suite, XSBench \cite{Tramm2014},
the Random access workload from the GUPS suite~\cite{Plimpton2006}, Sparse Length Sum from DLRM~\cite{dlmr} and kmer-count from GenomicsBench~\cite{genomicsbench}. 
\konkanellorevc{We extract the page size information for each workload from a real system that uses Transparent Huge Pages~\cite{corbet2011,arcangeli2010} with both 4KB and 2MB pages. 
The fraction of 2MB pages is shown in Table~\ref{tab:workloads}}.
We create five mixes of 2, 4, and 8 benchmarks to evaluate multi-programmed workloads.
Each benchmark is executed for 500M instructions.

\begin{table}[ht!]
  \centering
  \scriptsize
  \caption{Evaluated Workloads}
  \vspace{-2mm}
    \begin{tabular}{m{8em}m{22em}r}
    \toprule
    
    \textbf{Suite} & \textbf{Workload \konkanellorevc{(Fraction of 2MB pages~\cite{corbet2011})}} & \textbf{Input size} \\
        \midrule

    GraphBIG~\cite{Lifeng2015} & Betweeness Centrality (BC) \konkanellorevc{(36\%)}, Bread-first search (BFS) \konkanellorevc{(46\%)}, Connected components (CC) \konkanellorevc{(55\%)},
     Coloring (GC) \konkanellorevc{(52\%)}, PageRank (PR) \konkanellorevc{(51\%)}, Triangle counting (TC) \konkanellorevc{(32\%)}, Shortest-path (SP) \konkanellorevc{(46\%)}   & 8 GB \\
    % \cmidrule{2-3}          & Bread-first search (BFS) & 7.6 GB \\
    % \cmidrule{2-3}          & Connected components (CC) & 7.9 GB \\
    % \cmidrule{2-3}          & Graph coloring (GC) & 7.8 GB \\
    % \cmidrule{2-3}          & PageRank (PR) & 8.2 GB \\
    % \cmidrule{2-3}          & Triangle counting (TC) & 7.0 GB \\
    % \cmidrule{2-3}          & Shortest-path (SP) & 7.2 GB \\
    \midrule
    XSBench~\cite{Tramm2014} & Particle Simulation (XS) \konkanellorevc{(43\%)}      & 9 GB \\
    \midrule
    GUPS~\cite{Plimpton2006}  & Random-access (RND) \konkanellorevc{(51\%)}  & 10 GB \\
    \midrule
    DLRM~\cite{dlmr}  & Sparse-length sum (DLRM) \konkanellorevc{(46\%)}  & 10.3 GB \\
    \midrule
    GenomicsBench~\cite{genomicsbench} & k-mer counting (GEN) \konkanellorevc{(51\%)} & 33 GB \\
    \bottomrule
    \end{tabular}
  \label{tab:workloads}%
      \vspace{-2mm}
\end{table}%

\textbf{Evaluated Systems.} Table~\ref{tab:simconfig} shows the configuration of the simulated systems. We evaluate five different systems: 
(i) \textbf{Radix}: Baseline x86-64 system that uses (1) \konrevf{the conventional fully-flexible address mapping} and (2) \konrevw{a conventional} four-level radix-based page table.
(ii) \textbf{POM-TLB}: a system that employs a large 64K-entry software L3 TLB~\cite{pomtlbISCA2017} to increase TLB reach and reduce the number of PTWs.
(iii) \textbf{ECH}: a system that uses (1) the conventional fully-flexible address mapping and (2) the state-of-the-art hash-based page table, 
Elastic Cuckoo Hash Table~\cite{elastic-cuckoo-asplos20} (ECH). 
ECH employs $n$ different hash tables and issues $n$ memory requests in parallel to each one of the hash tables
to increase \konrevf{parallelism} and reduce PTW latency. ECH makes use of additional Cuckoo Walk Caches to avoid looking up all hash tables. 
We implement an optimistic version of ECH that (1) \konrevf{does not require allocating large contiguous physical memory blocks to store the page table} and (2) performs migrations \konrevf{between page tables} without \konrevw{any} performance \konrevw{penalty}.
Thus, we provide and upper bound estimate of ECH’s performance as described in~\cite{mehtJovanHPCA2023}. 
(iv) \textbf{RMM}: a system that uses multiple dynamically-allocated contiguous physical regions, called ranges~\cite{karakostas2015}, to provide efficient address translation for a small number of large memory objects used by the application. 
RMM uses a hardware Range Lookaside Buffer (RLB) to cache the mappings of the ranges and allocates ranges using a custom memory allocator~\cite{karakostas2015}.
(v) \textbf{Utopia:} a system that employs Utopia. Utopia uses both the page-fault-based allocation policy and the PTW-Tracking-based migration policy (\S\ref{sec:heuristics}). 
Utopia employs (i) two 512MB \utopiasegs, one for storing 4KB pages and one for storing 2MB pages and (ii) the rest of memory is organized as a \flexseg.  
(vi) \textbf{Perfect TLB}: a system where every address translation requests hits in a perfect L1 TLB (P-TLB). P-TLB provides an upper bound estimate of the performance gains \konrevf{possible} by accelerating address translation.

\konrevf{We provide additional 2MB pages (1GB in total) to Radix, ECH, and POM-TLB, to match the size of the contiguously allocated RestSegs and conduct a fair comparison against Utopia. For RMM, we provide an additional 1GB contiguous physical memory block.
In all evaluated systems, the L2 TLB access is performed in parallel with the PTW to conduct a fair comparison against Utopia.}

\section{Evaluation Results}
\label{sec:evaluation}

\subsection{Single-Core Results}
\label{sec:evaluation-sc}

\begin{figure}[ht!]
    \vspace{-2mm}
    \centering
    \includegraphics[width=\linewidth]{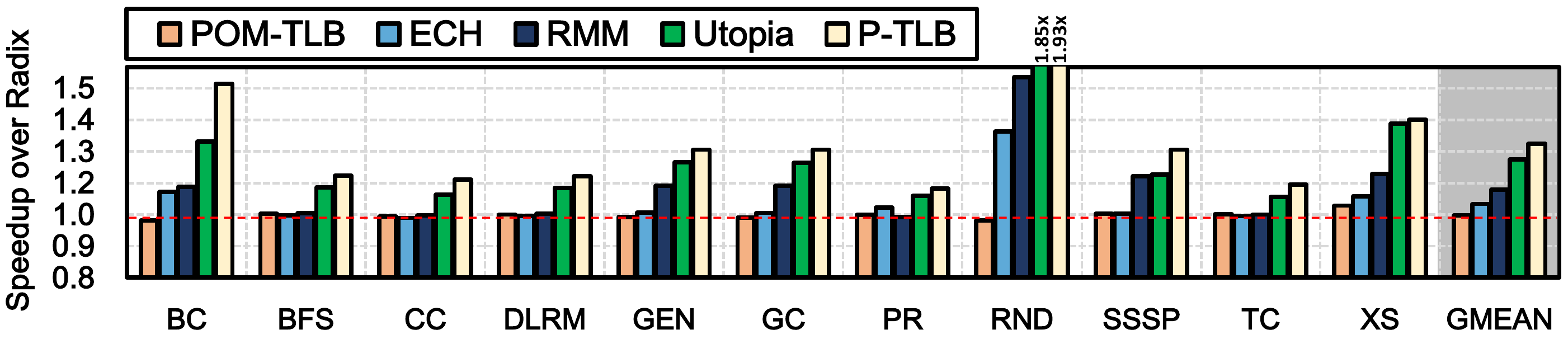}
    \vspace{-7mm}
    \caption{Speedup provided by POM-TLB, ECH, RMM, Utopia and P-TLB over Radix.}
    \label{fig:speedup_sc}
    \vspace{-3mm}

\end{figure}

Figure \ref{fig:speedup_sc} shows the execution time speedup of POM-TLB, ECH, RMM, Utopia and P-TLB compared to Radix, 
in \konrevf{the} single-core configuration, across 11 workloads. We make two key observations.  
First, Utopia on average outperforms Radix, POM-TLB, ECH, and RMM by $\speedupsc\%$, $\speedupoverpomtlb\%$, $\speedupoverech\%$, and $\speedupoverrmm\%$,  respectively.
Second, Utopia achieves \speedupwithinideal\% of the performance of the Perfect-TLB.
To better understand the performance speedup achieved by Utopia, we \konrevf{next} examine the impact of Utopia on
(i) address translation latency and (ii) translation-induced interference in main memory. 

\begin{figure}[h!]
    \vspace{-3mm}
    \centering
    \includegraphics[width=\linewidth]{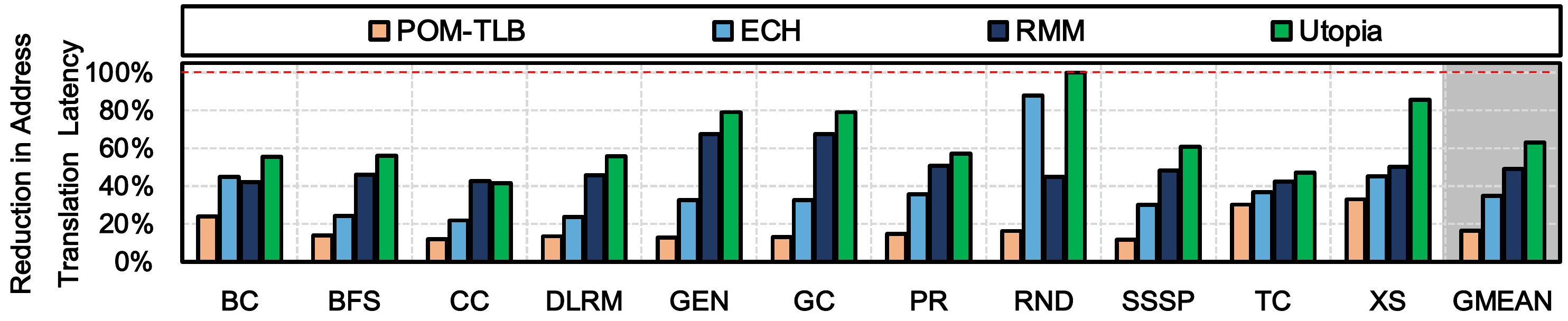}
    \vspace{-7mm}
    \caption{Reduction in address translation latency provided by POM-TLB, ECH, RMM and Utopia over Radix.}
    \label{fig:translate}
    \vspace{-2mm}
\end{figure}

 \head{Translation Latency} 
Figure \ref{fig:translate} shows the reduction in address translation latency provided by POM-TLB, ECH, RMM and Utopia over Radix.
We observe that Utopia significantly reduces address translation latency by 63\%, 47\%, 29\% and 14\% compared to Radix, POM-TLB, ECH, and RMM, respectively. 
This is because (i) RSWs are on average 7.6$\times$ faster than PTWs and (ii) Utopia reduces the number of PTWs by 78\% over Radix, on average across all workloads.

\begin{figure}[h!]
    \vspace{-3mm}
    \centering
    \includegraphics[width=\columnwidth]{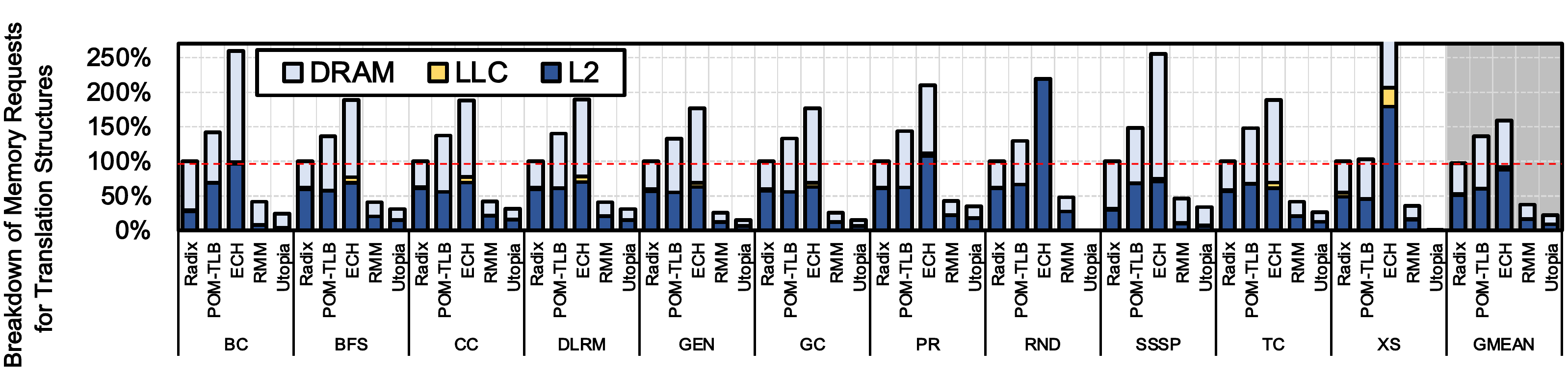}
    \vspace{-7mm}
    \caption{\konrevf{Breakdown of the \konrevb{servicing} location of memory requests that fetch translation metadata in POM-TLB, ECH, RMM and Utopia, normalized to Radix.}}
    \label{fig:translation_hit_eval}
    \vspace{-3mm}

\end{figure}

\konrevf{To better understand the \konrevw{sources} of Utopia's address translation efficiency, 
Figure~\ref{fig:translation_hit_eval} shows the breakdown of the servicing location (L2, LLC or DRAM) of
memory requests \konrevd{issued by the MMU to access the translation structures of}  POM-TLB, ECH, RMM and Utopia, normalized to Radix.}
We make three key observations. 
First, Utopia issues on average 88\% \konrevf{fewer} memory requests to the memory hierarchy compared to Radix.
Second, Utopia reduces the number of memory requests sent to DRAM, by 78\%, 82\%, 86\% and 40\% compared to Radix, POM-TLB, ECH and RMM, respectively. 
Third, 51\% of memory requests issued by Utopia (during a FSW or RSW) hit the L2 cache. 
We conclude that Utopia significantly reduces the number \konrevf{of memory requests issued to the memory hierarchy to access translation structures},
which leads to a significant reduction in address translation latency.
 %\footnote{Number of ranges is on average 20x higher in than the size of the 32-entry Range Lookaside Buffer which leads to 54\% RLB miss ratio}

\begin{figure}[h!]
    \centering
    \includegraphics[width=\linewidth]{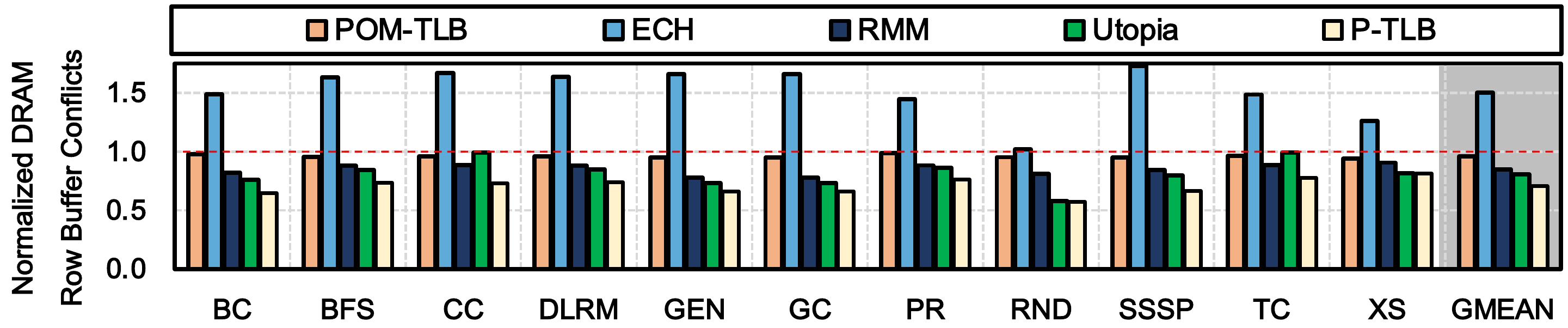}
    \vspace{-7mm}
    \caption{Reduction in DRAM row buffer conflicts provided by POM-TLB, ECH, RMM, Utopia and P-TLB over Radix.}
    \label{fig:rbmisseseval}
    \vspace{-3mm}
\end{figure}

\head{Memory Interference} Figure~\ref{fig:rbmisseseval} shows the reduction of DRAM row buffer conflicts (RBC) for POM-TLB, ECH, RMM, Utopia and P-TLB normalized to Radix. 
We make two key observations.
First, due to the reduced number of DRAM row activations to fetch translation metadata (Fig. \ref{fig:translation_hit_eval}),
Utopia reduces the \konrevf{total number of DRAM RBCs (i.e., considering row activations to access \konrevw{both} application data and translation structures)} by \konkanello{20\%, 15\%, 70\% and 4\%} over Radix, POM-TLB, ECH and RMM, respectively.
Second, Utopia causes only \konkanello{9\%} more DRAM RBCs compared to Perfect-TLB. 
We conclude that Utopia significantly reduces translation-induced interference in the main memory.

% \subsection{Comparison against Software-based TLB}
% We compare Utopia against POM-TLB~\cite{POTM}, a system that uses a TLB hierarchy with an L3 software-based TLB. For every L2 TLB miss, the MMU requests to the caches to lookup a 64K-entry software TLB.  Figure~\ref{fig:potmspeedup} shows the speedup of Utopia and POM-TLB compared to Radix. As we observe, POM-TLB outperforms Radix by 5\% while Utopia significantly outperforms POM-TLB by 19\%. Even though POM-TLB reduces the amount of PTWs by 39\%, it only leads to an average 25\% lower translation latency compared to Radix. The latency reduction is small because POM-TLB is designed for virtualized environments where the PTW latency is much higher compared to native environments. 

\subsection{Multi-Programmed Results}

\begin{figure}[h!]
    \vspace{-2mm}
    \centering
    \includegraphics[width=0.95\linewidth]{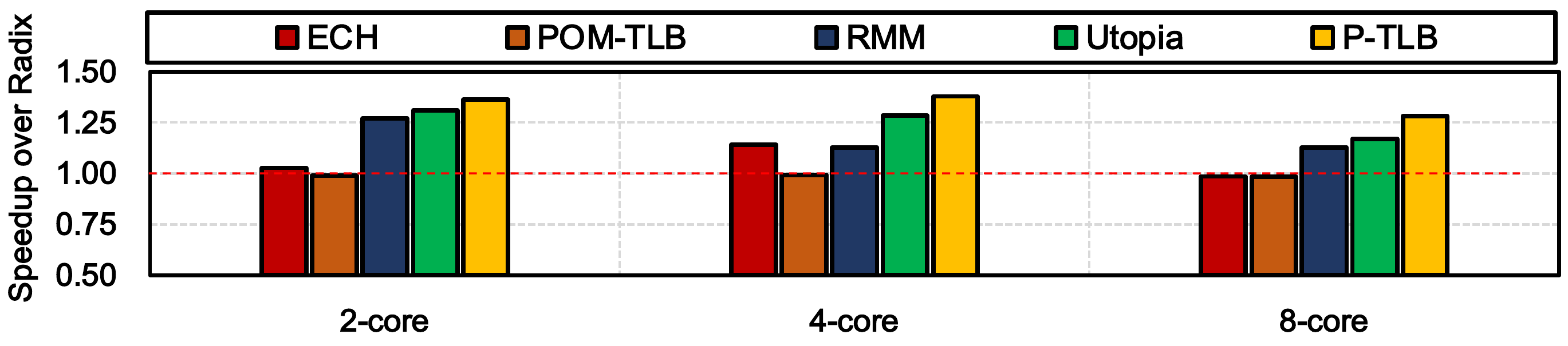}
    \vspace{-4mm}
    \caption{Speedup achieved by ECH, POM-TLB, RMM, Utopia compared to Radix across increasing core counts. \konrevf{Note that y-axis starts at 0.50.}}
    \label{fig:multiprog4812}
    \vspace{-2mm}

\end{figure}

We evaluate Utopia in 2-, 4- and 8-core systems using multi-progra\-mmed workloads to demonstrate 
its effectiveness in systems where multiple applications compete for memory space in the same RestSeg.
Figure \ref{fig:multiprog4812} shows the average (of 5 mixes) performance speedup of ECH, POM-TLB, RMM, Utopia and P-TLB compared to Radix. 
First, we observe that Utopia outperforms Radix, POM-TLB, ECH and RMM across all \konrevf{multi-core} systems. 
In the 8-core system, Utopia outperforms the second-best performing mechanism (RMM) by 5\% and achieves 91\% 
of the performance of P-TLB. We conclude that Utopia \konrevf{provides} high performance \konrevf{benefits},
\konrevf{even when} multiple applications compete for memory space in the same RestSeg.

\subsection{\konrevf{Analysis of} Utopia}

\subsubsection{\konrevw{Effectiveness of TAR and SF Caches.}} 
\begin{figure}[ht!]
     \vspace{-2mm}
    \centering
    \includegraphics[width=\columnwidth]{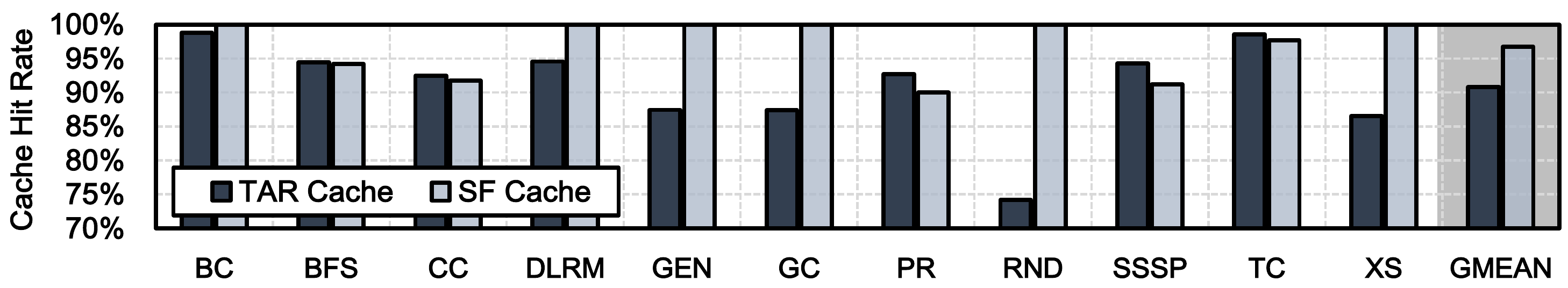}
    \vspace{-7mm}
    \caption{Hit-rate of TAR/SF cache.}
    \label{fig:sftarmig}
    %\vspace{-1mm}
    \vspace{-5mm}
\end{figure}

To better understand why RSWs are 
more efficient than PTWs, we examine the effectiveness of the TAR/SF caches.
Figure~\ref{fig:sftarmig} demonstrates the hit rate \konrevf{of the} TAR cache and the SF cache.
\konrevf{We observe that 81\% of TAR requests hit in the TAR cache while 98\% of SF requests hit in the SF cache.
This is because TAR/SF entries experience high reuse (both data structures consume 545KB in total for a 512MB RestSeg). 
TAR cache has a lower hit rate compared to SF cache because TAR is 31x larger than SF (i.e., 528KB vs 17KB) and the \konrevw{2KB TAR} cache 
cannot cover as many TAR entries as the SF cache does. 
We conclude that TAR/SF caches are highly effective and lead to low-latency RSWs.}

\subsubsection{Effect of Utopia's Page Migrations.}
\begin{figure}[h]
        \vspace{-2mm}
        \centering
        \includegraphics[width=\columnwidth]{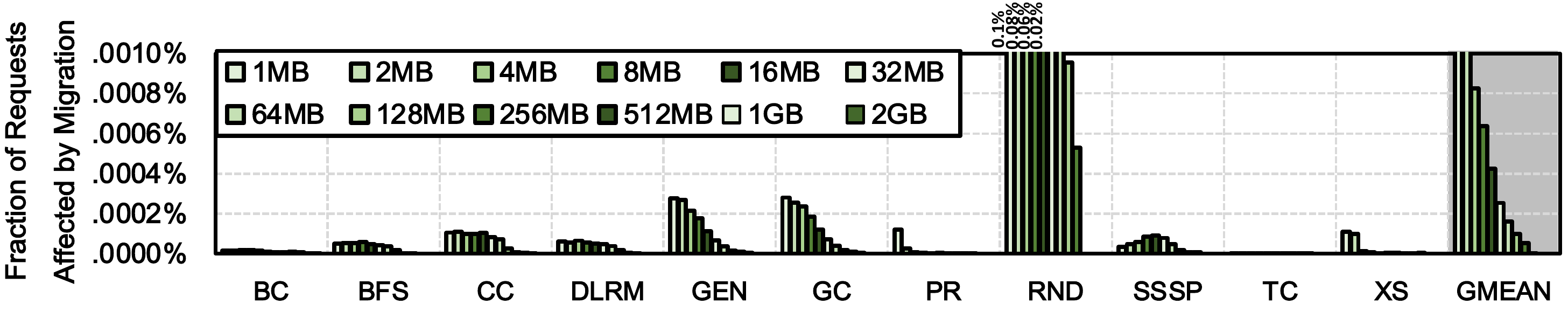}
        \vspace{-8mm}
        \caption{\konkanelloreva{Fraction of memory requests that get stalled \konrevf{due to} migrations, across RestSegs with different sizes.}}
        \label{fig:migoverhead}
        \vspace{-5mm}
\end{figure}

To better understand how page migration affects memory requests (\S\ref{sec:os-support}), 
\konrevf{we plot (in Figure~\ref{fig:migoverhead})} the fraction of memory requests that get stalled by page migrations across RestSegs with different sizes. 
We observe that less than 0.001\% of the memory requests \konrevf{are affected due to migration}, even for the smallest RestSeg (i.e., 1MB).
\konrevf{This is due to two reasons. First, the number of \konrevw{page} migrations is low (i.e, 0.8 migrations per kilo instructions on average).
Second, 82\% of the  page migrations occur due to evictions  from a \utopiaseg to a \flexseg.
As we show in Table~\ref{tab:simconfig}, Utopia chooses pages with low reuse for eviction (and eventually migration) candidates (i.e., SRRIP replacement policy~\cite{srrip}). Thus, 
the probability of accessing \konrevw{a} migrated page \konrevw{is low, especially} during the migration \konrevw{process}.
\konrevw{We conclude \konrevx{that} Utopia's page migrations minimally interfere with regular memory requests.}}

% \head{RestSeg Utilization}
% To understand the impact of Utopia to memory utilization we evaluate the utilization of the RestSeg across time. Figure~\ref{fig:restsegutil} shows the memory utilization of the 4KB RestSeg across 500 one-million instruction epochs for five workloads. As we observe, the RestSeg gets filled up completely after a few epochs due to the page-fault-based allocation policy. Hence, we conclude that RestSegs do not suffer from memory underutilization.

% \begin{figure}[ht!]
%      \vspace{-2mm}
%     \centering
%     \includegraphics[scale=0.4]{figures/restsegutil.pdf}
%     \vspace{-4mm}
%     \caption{Utilization of RestSeg across 1M-instr epochs.}
%     \label{fig:restsegutil}
%     \vspace{-2mm}
% \end{figure}

\subsubsection{\konrevw{Utopia's Effectiveness in Discovering Costly-to-Translate Pages.}} 
\begin{figure}[h!]
    \centering
    \vspace{-2mm}
    \includegraphics[width=\linewidth]{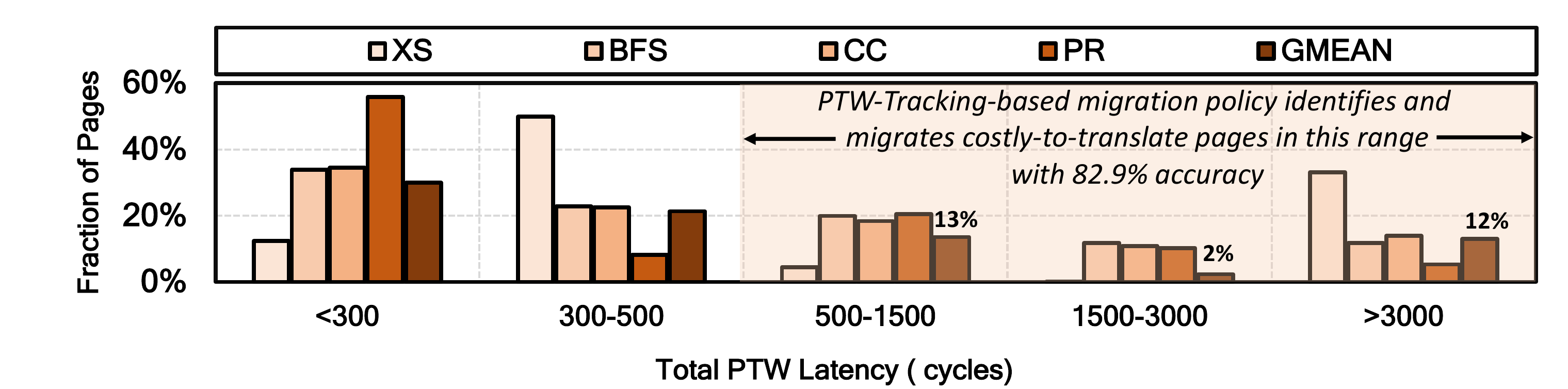}
    \vspace{-7mm}
    \caption{\konrevw{Distribution of PTW latencies across pages in Radix.}}
    \label{fig:histoptw}
    \vspace{-3mm}
\end{figure}

\konkanelloreve{To understand the \konrevw{effectiveness of} Utopia in discovering costly-to-translate pages, Figure~\ref{fig:histoptw} plots the fraction of pages that experience different PTW latencies for four workloads \konrevw{in the baseline system \konrevx{Radix}}:
(i) less than 300 cycles, (ii) 300-500 cycles, (ii) 500-1500, (iii) 1500-3000 and (iv) more than 3000.
\konrevw{We observe that} more than 50\% of pages experience latencies between 500 and 3000, while more than 25\% of pages experience latencies larger than 3000 cycles.
The PTW-Tracking-based migration policy (\S\ref{sec:heuristics}) identifies and migrates to RestSeg costly-to-translate pages (>500-cycle PTW latency) with 82.9\% accuracy. 
We conclude that the evaluated workloads have a significant number of costly-to-translate pages which Utopia can effectively identify 
and migrate to a RestSeg and \konrevw{thus} reduce \konrevw{the} address translation latency.}

\subsubsection{Effectiveness of RestSeg's Replacement Policy.}

\begin{figure}[ht]
    \centering
    \includegraphics[width=\linewidth]{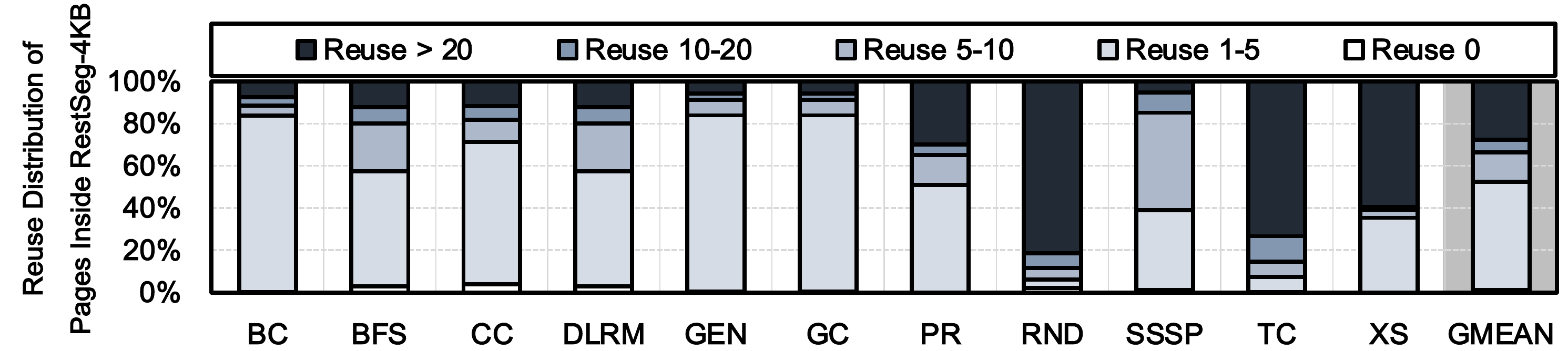}
    \vspace{-7mm}
    \caption{Reuse-level distribution of (4KB) pages that reside in the RestSeg.}
    \label{fig:restsegreuse}
    \vspace{-3mm}
\end{figure}

Figure~\ref{fig:restsegreuse} shows the reuse-level distribution of \konrevw{(4KB)} pages while they reside in the RestSeg.
\konrevx{This reuse level corresponds to how many times RSWs resolve address translation before evicting the page from the RestSeg}.
\konrevx{We make two key observations. First, nearly 0\% of the pages in the RestSeg are evicted without being reused, which indicates that 
Utopia saves at least one PTW for almost every page that is allocated or migrated to a RestSeg.}
Second, we observe that more than 50\% of pages experience reuse \konrevx{larger} than 5
and 27\% of the pages experience reuse \konrevx{larger} than 20 before getting evicted from the RestSeg. 
This \konrevx{relatively high reuse} is due to two reasons. First, the PTW-Tracking-based migration policy (\S\ref{sec:heuristics}) migrates pages that experience highly-frequent PTWs to a RestSeg, \konrevx{which converts the slow PTWs to fast RSWs}.
Second, the SRRIP replacement policy~\cite{srrip} (employed by the RestSeg) \konrevx{effectively} estimates the re-reference interval (analogous to reuse) of pages and evicts pages with low reuse. 
We conclude that Utopia using (i) the PTW-Tracking-based migration policy and (ii) the SRRIP replacement policy in the RestSeg,
retains pages with high reuse in the RestSeg which leads to (i) fast address translation for pages that otherwise experience frequent PTWs in the baseline system and (ii)
efficient utilization of the RestSeg.

% \begin{figure}[ht]
%     \centering
%     \includegraphics[width=\linewidth]{figures/micro2023_figures/reuse_restseg.pdf}
%     \vspace{-7mm}
%     \caption{Breakdown of translation reuse (i.e., how many times translation was resolved by an RSW before a page is migrated to a FlexSeg) of pages which were migrated using the page table-based migration policy.}
%     \label{fig:restsegreuse}
%     \vspace{-2mm}
% \end{figure}

\subsubsection{Sensitivity to \utopiaseg Size.}
\begin{figure}[h!]
     \vspace{-2mm}
    \centering
    \includegraphics[width=\columnwidth]{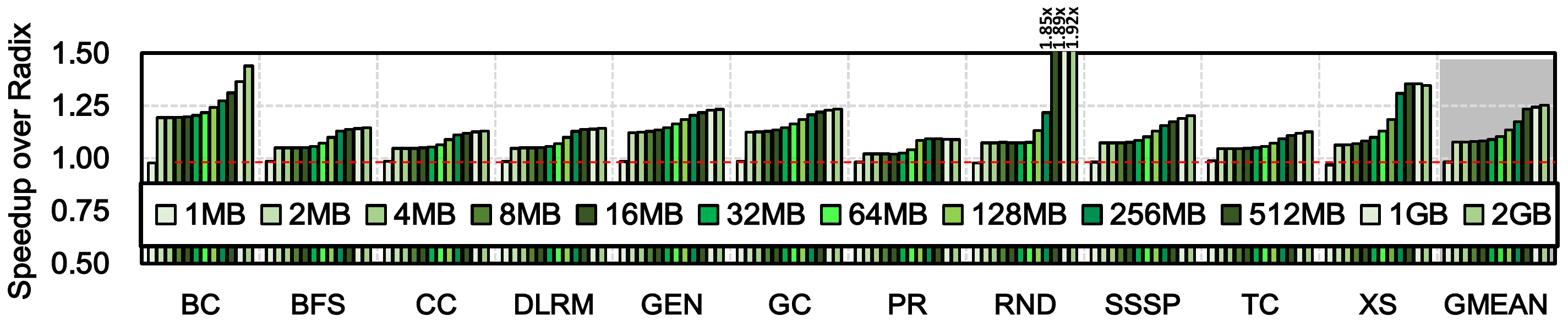}
    \vspace{-7mm}
    \caption{\konkanellorevb{Speedup provided by Utopia over Radix, across different RestSeg sizes.}}
    \label{fig:sens_size}
    \vspace{-3mm}

\end{figure}

\konkanellorevb{Fig.~\ref{fig:sens_size} shows the execution time speedup provided by Utopia over Radix across different \utopiaseg sizes. 
We make three key observations. 
First, larger \utopiasegs lead to higher performance benefits, up to 27\% for the 2GB \utopiaseg. 
Second, the performance of the 512MB \utopiaseg is within 1.3\% of the performance achieved by the 2GB \utopiaseg. 
Third, the 1MB RestSeg achieves the same performance as the baseline configuration, since it does \emph{not} provide enough space to store costly-to-translate pages.
We chose the 512MB \utopiaseg in our evaluation setup since it delivers similar performance gains to the 2GB \utopiaseg with $4\times$ lower memory consumption.}

\subsubsection{Effect of Performing RSW in Parallel to L2 TLB Access.}
\begin{figure}[h!]
    \vspace{-2mm}
    \centering
    \includegraphics[width=\linewidth]{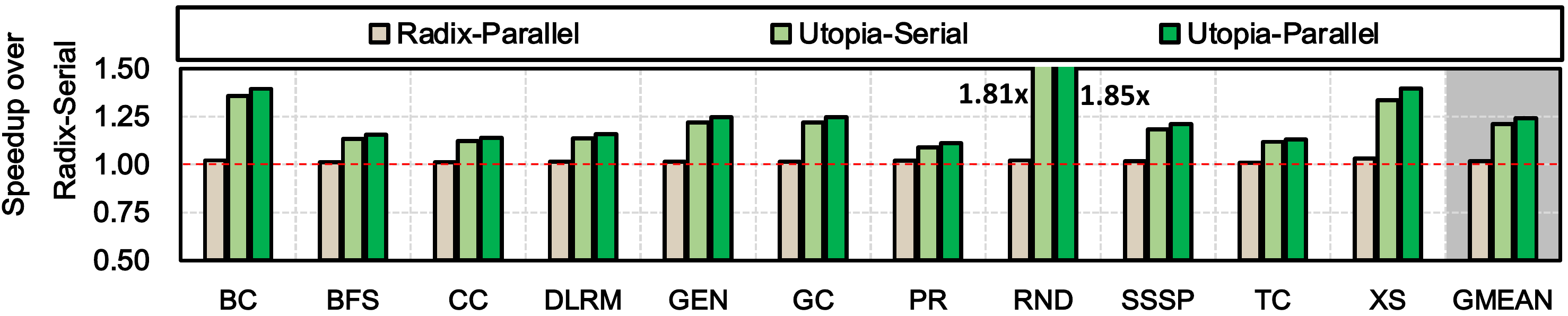}
    \vspace{-7mm}
    \caption{\konrevw{Speedup provided by Utopia with parallelized RSW/L2 TLB access and Radix with parallelized PTW/L2 TLB access, compared to Radix with serialized PTW/L2 TLB access.}}
    \label{fig:serial}
    \vspace{-2mm}

\end{figure}

To understand the benefits of performing the RSW in parallel with the L2 TLB access, Figure~\ref{fig:serial} shows the execution time speedup
of: \konrevl{1) Radix that performs the PTW in parallel with the L2 TLB access  (Radix-Parallel) and 2) Utopia that performs the RSW in serial/parallel with the L2 TLB access (Utopia-Serial/Utopia-Parallel), 
\konrevw{all normalized} to the performance of Radix that performs the PTW in serial with the L2 TLB access (Radix-Serial)}.
We make two key observations. First, Utopia-Serial outperforms Radix-Serial by 21\% on average across all workloads.
Second, Utopia-Parallel outperforms Utopia-Serial by 3\% on average. 
We conclude that Utopia (i) \konrevx{outperforms Radix in both serial and parallel L2 TLB/RSW configurations} and 
(ii) parallelizing the RSW with the L2 TLB access provides a noticeable performance benefit compared to serializing the RSW with the L2 TLB access.

\subsubsection{Effect of Utopia on Non-Translation-bound Workloads.}
\begin{figure}[h!]
    \vspace{-2mm}
    \centering
    \includegraphics[width=\linewidth]{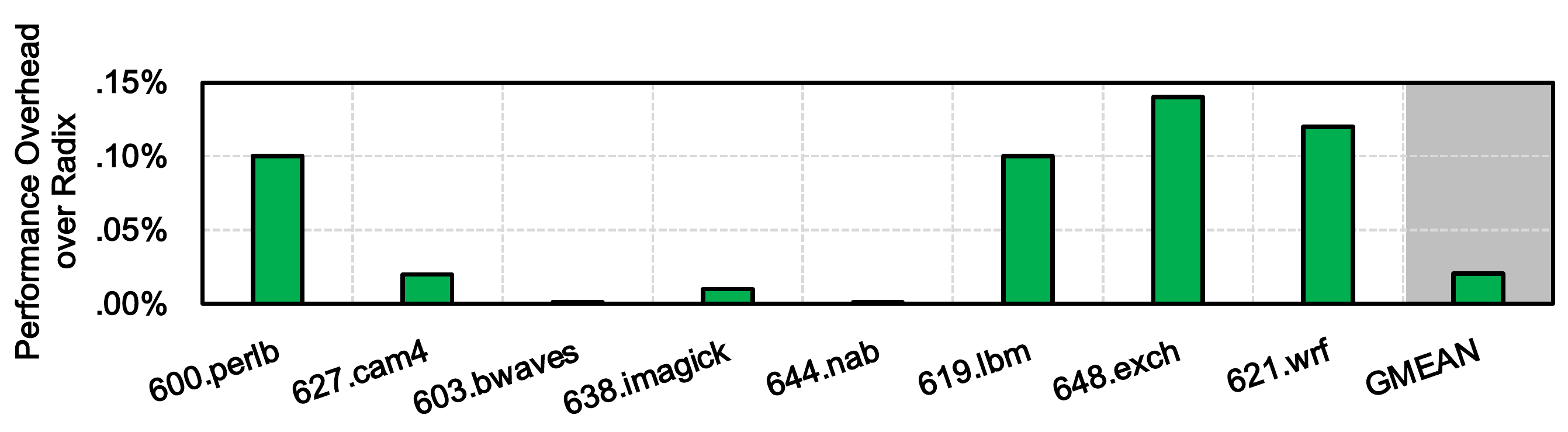}
    \vspace{-7mm}
    \caption{{Performance overhead of Utopia compared to Radix in non-translation-intensive workloads.}}
    \label{fig:spec}
    \vspace{-3mm}

\end{figure}

{To understand the performance impact of Utopia on workloads that do not experience high performance overheads due to address translation, we evaluate 
Utopia and Radix 
using eight workloads from the SPEC CPU2017 benchmark suite~\cite{spec2017} that exhibit low (i.e., smaller than 2) L2 TLB MPKI. 
Figure~\ref{fig:spec} shows the performance \konrevw{loss of Utopia compared to Radix.} 
We observe that Utopia causes less than 0.05\% \konrevw{average performance loss}.
This \konrevw{very low overhead} is due to two factors: (i) non-translation-bound workloads experience high L1/L2 TLB hit rates (more than 95\% on average across all workloads)
and Utopia does not affect the L1/L2 TLB \konrevw{hit rate or latency} and 
(ii) Utopia does not trigger many migrations from a FlexSeg to a RestSeg (i.e., most of the pages \konrevw{are correctly estimated to be} \konrevw{not} costly-to-translate) 
and \konrevw{thus} avoids \konrevw{evicting} high-locality data from the cache hierarchy. 
We conclude that Utopia \konrevw{has negligible} performance \konrevw{impact} to applications that do not experience high translation overheads.}

\subsubsection{Sensitivity to RestSeg Address Mapping Function.}
\begin{figure}[h!]
     \vspace{-2mm}
    \centering
    \includegraphics[width=\columnwidth]{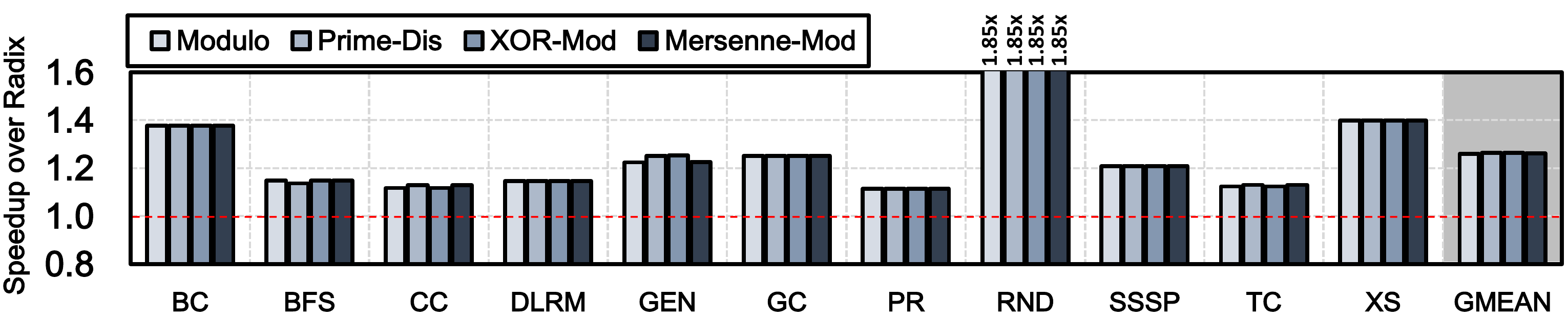}
    \vspace{-7mm}
    \caption{Speedup provided by Utopia with \konrevw{different RestSeg address mapping functions}.}
    \label{fig:hashfunc}
    \vspace{-3mm}
\end{figure}

To understand the impact of the \konrevw{address mapping function} that determines the location of a page inside a RestSeg, Figure~\ref{fig:hashfunc} plots the performance of 
Utopia using four different hash functions: (i) modulo hashing (MOD), (ii) prime displacement hashing~\cite{prime-disHPCA2004}, (iii) XOR-based hashing~\cite{xorbasedComp2008} and (iv) Mersenne modulo hashing~\cite{mersenne}. 
We observe that the modulo hash function performs similarly to more sophisticated hash functions while \konrevw{requiring minimal hardware support.}
\konrevw{As such}, the modulo hash function provides the best performance/complexity trade-off and we use it in our evaluation setup.

\subsubsection{Sensitivity to Context Switches.} To evaluate the impact of context switches on Utopia's performance, we evaluate 
Utopia and Radix using different context switch quanta (CSQ) while multiple workloads execute in a single core in a round-robin manner.
Each workload is executed for CSQ time units before the OS performs a context switch to the next workload.

\begin{figure}[h!]
    \vspace{-2mm}
    \centering
    \includegraphics[width=\linewidth]{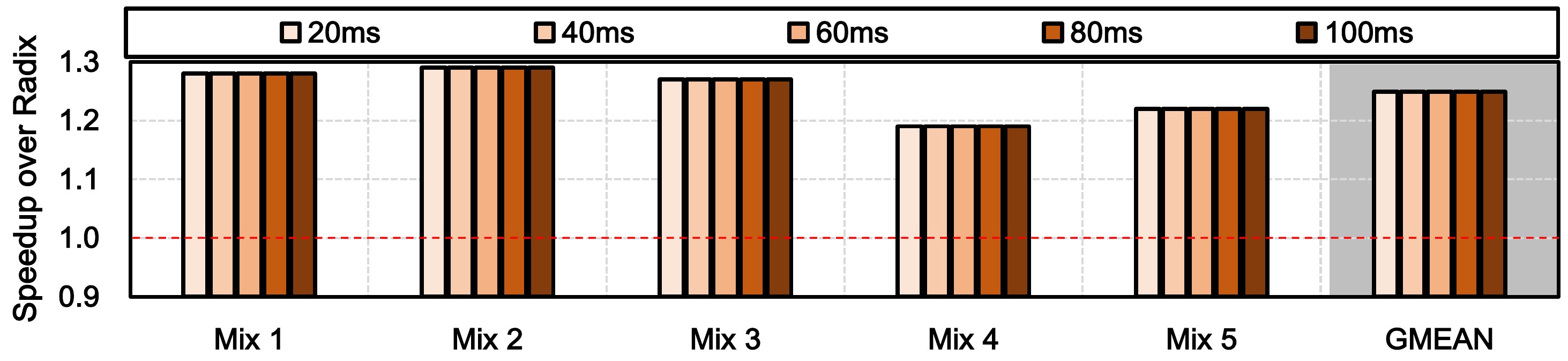}
    \vspace{-7mm}
    \caption{Speedup provided by Utopia over Radix while executing multiple applications in a single core for different context switch quanta values.}
    \label{fig:context}
    \vspace{-2mm}
\end{figure}

Figure~\ref{fig:context} shows the performance of Utopia and Radix using 5 different CSQs, ranging from 20ms to 100ms, across 5 mixes of 5 different workloads.
We make two observations.
First, Utopia \konrevw{provides} on average 24.9\% higher performance compared to the baseline system for the smallest CSQ \konrevw{(i.e., 20~ms)}, across all 5 workload mixes. 
Second, we observe that increasing the CSQ does not affect the performance of Utopia. 
This is because Utopia does not affect context switch overhead and Utopia's 
operations are not affected by the CSQ.

\section{Related Work}

To our knowledge, Utopia is the first hybrid virtual-to-physical address mapping \konrevw{technique} that enables \emph{both} flexible and restrictive hash-based virtual-to-physical address mapping schemes to harmoniously \emph{co-exist} in the system.
We already comprehensively compared Utopia to systems that employ (i) large software-managed TLBs~\cite{pomtlbISCA2017}, 
(ii) state-of-the-art page table designs~\cite{elastic-cuckoo-asplos20} and (iii) contiguity-aware translation mechanisms~\cite{karakostas2015} in \S\ref{sec:evaluation-sc}.
In this section, we qualitatively compare Utopia to other related prior works that propose solutions to reduce address translation overheads.

\head{Hash-based Address Mapping} \konrevw{Multiple} works~\cite{nearmemoryPact17,smith,mosaicpagesASPLOS2023} leverage the concept of hash-based address mapping to accelerate address translation. 
%Both works focus on organizing the whole main memory with a restrictive hash-based mapping  that fundamentally limits the flexibility of virtual memory. 
Picorel et al. \cite{nearmemoryPact17} and Gosakan et al.~\cite{mosaicpagesASPLOS2023} propose employing a restrictive hash-based virtual-to-physical address mapping across the \konrevw{entire} main memory to reduce the overheads of address translation.
Although such techniques drastically reduce address translation overheads, they generally handicap core VM features such as (i) sharing pages and (ii) the flexibility of allocating pages in free memory space to avoid swapping.
For example, as we show in \S\ref{sec:motivation}, the mapping scheme proposed by Picorel et al.~\cite{nearmemoryPact17} leads to a \konrevw{large} increase ($2.2\times$ on average) 
in accesses to the swap space over the baseline system.
In contrast, Utopia accelerates address translation while supporting all the key features enabled by the conventional virtual memory framework.

\head{Efficient TLBs and Page Walk Caches (PWCs)} Many prior works focus on reducing address translation overheads through efficient TLB and PWC designs\VMtlball.
Such techniques involve: (i) prefetching TLB and page table entries\VMtlbprefetching, 
(ii) TLB-specific replacement policies\VMtlbreplacementpolicy, (iii) employing software-managed TLBs\VMsoftwareTLB,
(iv) sharing TLBs across cores\VMtlblthree, (v) employing efficient PWCs\VMpwcs, and (vi) \konrevs{PT-aware cache management~\cite{pinningAccess2022,flataAsplos2022,consciousISPASS2022} (e.g., specialized cache replacement policies for PTEs~\cite{consciousISPASS2022}).}
% For example, Vavouliotis et al.~\cite{vavouliotis2021} propose a composite TLB prefetcher that prefetches TLB entries based on the locality of PTEs. 
% Margaritov et al.~\cite{margaritov2019prefetched} propose a page table prefetching scheme to enable direct accesses to the last levels of the page table.
% Reyoo et al.~\cite{pomtlbISCA2017} propose using a software-managed L3 TLB to back up the last-level 
% TLB and reduce PTWs in virtualized environments.
Although such techniques may offer notable performance improvements, \konrevw{their effectiveness \konrevf{reduces} as the page table size increases}.
This is because they rely on (i) the existing scarce TLB resources, which are unable to accommodate the \konrevw{very} large number of virtual-to-physical mappings required by data-intensive applications or (ii) 
new hardware/software translation structures that pose a significant trade-off between performance and area/power/energy efficiency.
In contrast, Utopia  fundamentally reduces the size of the \konrevw{address translation} structures to enable efficient address translation. 
However, we believe that Utopia can be used in combination with techniques that \konrevw{improve the efficiency of the} TLB hierarchy and PWCs, 
to further reduce address translation overheads. For example, caching TLB entries in the L2 cache~\cite{kanellopoulosMICRO2023victima}, \konrevw{prefetching TLB and PT entries\VMtlbprefetching}, and
employing efficient PWCs\VMpwcs~can significantly reduce \konrevw{the} L2 TLB miss latency (when \konrevw{the needed data} does not reside in a RestSeg)
and further accelerate address translation in Utopia.
% In contrast, Victima repurposes the \emph{existing} underutilized resources of the cache hierarchy to drastically increase 
% \konrevf{address} translation reach and thus does \emph{not} require additional structures to store translation metadata.
% For example, as we show in \S\ref{sec:motivation-stlb}, employing a software-managed TLB to back up the L2 TLB is not effective 
% in native environments as the latency of the PTW is similar to the latency of accessing the software-managed TLB. 
% In \S\ref{sec:virtualized_results} and \S\ref{sec:native_results}, we compare Victima against state-of-the-art software-managed TLB, POM-TLB~\cite{pomtlbISCA2017} and show that Victima
% outperforms POM-TLB by \speedupoverpomtlb\% (\speedupoverpomtlbvirt\%) in native (virtualized) environments \konrevs{by} storing TLB entries in the 
% high-capacity and low-latency L2 cache.

\head{Alternative Page Table Designs}
Various prior works propose (i) hash-based and (ii) flattened PT designs~to reduce PTW latency\VMpagetable.
For example, Skarlatos et al.~\cite{elastic-cuckoo-asplos20} and Stojkovic et al.~\cite{mehtJovanHPCA2023} propose replacing the radix-tree-based page table with a Cuckoo hash table~\cite{fotakis} to parallelize accesses 
to the PT and reduce PTW latency. Park et al.~\cite{flataAsplos2022} propose a flat PT design in combination with a PT-aware
replacement policy to \konrevs{reduce PTW latency}. In \S\ref{sec:evaluation}, we show that Utopia significantly outperforms a standalone version of ECH (i.e., in a system that employs the fully-flexible virtual-to-physical address mapping).
We believe that Utopia can be used in combination with alternative PT designs to further reduce the latency of FSWs \konrevw{(flexible segment walks)} and further accelerate address translation.

\head{Employing Large Pages} 
Many works propose hardware and software mechanisms for efficient support for pages of varying sizes\VMlargepages.  
For example, Panwar et al.\cite{panwar2019hawkeye,panwar2018making} propose new OS techniques to improve the efficiency of mechanisms that enable \konrevw{application-transparent} allocation of large pages.
As we discuss in \S\ref{sec:structure}, Utopia is backward compatible with large page mechanisms and the OS can create multiple RestSegs to accommodate pages of different sizes.

\head{Contiguity-Aware Address Translation} 
\konrevs{Many prior works} \konrevs{enable and exploit} \konrevs{virtual-to-physical address} contiguity to perform \konrevs{low-latency} address translation\VMcontiguity. 
For example, in \cite{vm2}, \konrevs{the authors propose pre-allocating arbitrarily-large contiguous physical regions (10-100's of GBs) to drastically increase the translation reach for specific data structures of the application.}
Karakostas et al. \cite{karakostas2015} propose the use of multiple dynamically-allocated contiguous physical regions, called ranges, to provide efficient address translation for a small number of large memory objects used by the application.
Although these works can significantly increase translation reach and reduce address translation overheads, their effectiveness heavily depends on the availability of free contiguous memory blocks.
Utopia allocates RestSegs during system boot \konrevw{time} to avoid \konrevw{the need to find or create} free contiguous memory blocks during runtime.
In \S\ref{sec:evaluation}, we show that, given the same amount of available contiguity in the system, Utopia outperforms (by \speedupoverrmm\%) a system that employs multiple segments with contiguous virtual-to-physical address mappings~\cite{karakostas2015}.
We believe that Utopia can be naturally extended to incorporate segments with contiguous virtual-to-physical address mappings to further reduce address translation overheads.

\head{Address Translation in Virtualized Environments}
\konrevs{Various works propose techniques to} reduce address translation overheads in virtualized environments\VMvirtualized.
For example, Ghandi et al. \cite{vm25} propose a hybrid address translation design for virtualized environments that combines shadow paging and nested paging.
Utopia can be employed to support and accelerate address translation in virtualized environments by placing costly-to-translate guest-virtual \konrevw{as well as} host-virtual pages in RestSegs.
We leave the evaluation of Utopia in virtualized environments as future work.

\head{Virtual Caching \& Intermediate Address Spaces} Another class of works focuses on delaying address translation by using techniques such as 
virtual caching\VMvirtualcaching~and intermediate address spaces\VMintermediate. 
Virtually-indexed caches reduce address translation overheads by performing address translation only after a memory request 
misses in the LLC~\cite{basu2012, cekleov1997a, cekleov1997b, wood1986}. 
\konrevw{Gupta et al.~\cite{midgard} propose mapping large virtual memory \konrevw{regions} to an intermediate address space to enable fast virtual-to-intermediate address translation and delay
intermediate-to-physical address translation until an LLC miss (for the corresponding data access)}. 
Hajinazar et al.~\cite{vbi} propose the use of virtual blocks to enable fast virtual-to-intermediate address translation and extend 
the memory controller with \konrevl{a programmable core} \konrevl{that handles (i) data allocation in physical memory and
(ii) intermediate-to-physical address translation}, based on properties of each virtual block.
Utopia is orthogonal to these techniques and can be used to accelerate address translation of costly-to-translate pages that reside either in the intermediate address space or in the virtual address space, regardless
of when \konrevw{the address translation} takes place (e.g., after an LLC miss or before the L1 cache access).

\vspace{-2mm}
\section{Conclusion}

We propose \emph{Utopia}, the first hybrid address mapping \konrevw{technique} that allows both flexible and restrictive address 
mapping schemes to harmoniously co-exist in \konrevx{a} system \konrevx{with virtual memory}. 
\konrevx{By restricting the virtual-to-physical address mapping, \papername alleviates the need to store a large number of virtual-to-physical mappings and 
 enables faster address translation via compact translation structures.}
At the same time, Utopia retains the ability to use the flexible address mapping to support conventional virtual memory features.
Our extensive evaluations using data-intensive workloads show that Utopia leads to fast and efficient address translation, \konrevw{improving application performance}  \konrevw{in both} single-core and multi-core systems. 
We believe that Utopia \konrevw{is also} applicable \konrevw{to} virtualized environments, address translation in GPUs and specialized accelerators, and \konrevw{address translation for I/O data}.
\konrevx{To enable further research in these and other directions, we open source Utopia at \textcolor{blue}{\url{https://github.com/CMU-SAFARI/Utopia}}.}
\vspace{-1mm}

%We propose \emph{Utopia}, the first virtual-to-physical address mapping mechanism that allow of flexible and restrictive address mapping schemes to co-exist in a general-purpose system. By mapping data to a segment that uses the restrictive address mapping, Utopia enables significantly lower address translation overhead whenever flexible address mapping is not necessary.
%Our evaluation results show that Utopia leads to efficient address translation in single-core, multi-core and virtualized environments. 

% We introduce Utopia, a new end-to-end translation subsystem which acts as a fast translation lane for frequently accessed but hard-to-translate pages (i.e., pages that the TLB subsystem does not efficiently translate). The key idea of Utopia is to track hard-to-translate pages and migrate them to a new set-associative physical segment, which we call UTopia Region (or for short, UTR), that enables a faster translation path to avoid page-table overheads, and is available to \emph{all} the processes in the system. Utopia comprises three lightweight key mechanisms to support the use of UTRs: (i) an application-transparent scheme to monitor hard-to-translate pages, (ii) architectural support to enable efficient discovery of the permissions and the location of pages which reside in a UTR and (iii) OS modules to handle page migration and replacement in a UTR. Our results show that Utopia increases the performance of memory-hungry single-core workloads by up to 1.8x and multi-programmed workloads by up to 1.85x, over a conventional TLB-based system.

\begin{acks}
  {
    We thank Jisung Park and Nika Mansouri-Ghiasi for their
    valuable feedback on this work. We thank Konstantinos Sgouras for his valuable help with 
    the development of the simulation infrastructure.
    We thank the anonymous reviewers of ISCA 2022, MICRO 2022, ISCA 2023, and MICRO 2023 for their feedback. We
    thank the SAFARI Research Group members for providing a
    stimulating intellectual environment. We acknowledge the generous gifts from our industrial partners: Google, Huawei, Intel,
    Microsoft, and VMware. This work is supported in part by
    the Semiconductor Research Corporation and the ETH Future
    Computing Laboratory. 
  }
\end{acks}

\bibliographystyle{ACM-Reference-Format}
\balance
% \clearpage
\bibliography{updated_file}

\end{document}